\documentstyle[12pt,aps]{revtex}

\begin{document} \draft

\preprint{HEP/123-qed}

\title{Self-organized criticality within generalized Lorenz scheme}

\author{Alexander I. Olemskoi
\footnote{Present address: Sumy State University,
Rimskii-Korsakov St. 2, 40007 Sumy, Ukraine;
electronic address: olemskoi@ssu.sumy.ua}}

\address{\it Max-Planck-Institut f\"ur Physik komplexer Systeme,
N\"othnitzer Strasse 38, D-01187 Dresden, Germany}

\author{Alexei V. Khomenko
\footnote{Electronic address: khom@phe.ssu.sumy.ua}}

\address{Sumy State University,
Rimskii-Korsakov St. 2, 40007 Sumy, Ukraine}

\author{Dmitrii O. Kharchenko
\footnote{Present address: Sumy State University,
Rimskii-Korsakov St. 2, 40007 Sumy, Ukraine;
electronic address: dikh@ssu.sumy.ua}
}
\address{\it Max-Planck-Institut f\"ur Physik komplexer Systeme,
N\"othnitzer Strasse 38, D-01187 Dresden, Germany}

\date{\today}

\maketitle

\begin{abstract}
The theory of a flux steady-state related to avalanche formation
is presented for the simplest model of a sand pile
within framework of the Lorenz approach.
The stationary values of sand velocity and
sand pile slope are derived as functions
of control parameter (driven sand pile slope).
The additive noises of above values are introduced to build the phase diagram,
where the noise intensities determine both avalanche and non-avalanche domains,
as well as mixed one.
Being corresponded to the SOC regime, the last domain is crucial
to affect of the noise intensities of vertical component
of sand velocity and sand pile slope especially.
To address to a self-similar behavior, a fractional feedback is used
as efficient ingredient of the modified Lorenz system.
In a spirit of Edwards paradigm, an effective thermodynamics is introduced
to determine a distribution over avalanche ensemble with negative temperature.
Steady-state behavior of the moving grains number,
as well as nonextensive values of entropy and energy is studied in detail.
The power-law distribution over avalanche sizes is
described within a fractional
Lorenz scheme, where noise of the energy plays a crucial role.
This distribution is shown to be solution of both fractional
Fokker-Planck equation and
nonlinear one. As a result, we obtain new relations between
exponent of the size distribution,
fractal dimension of phase space,
characteristic exponent of multiplicative noise, number of governing equations,
dynamical exponents and nonextensivity parameter.

\end{abstract}


\thispagestyle{empty}

\section{Introduction}\label{sec:level1}

In recent years considerable study has been given to the theory
of self-organized criticality (SOC)
that explains spontaneous (avalanche-type) dynamics,
unlike the typical phase transitions that occur
only when a control parameter is driven to a critical value \cite{0}, \cite{1x}.
A main feature of the systems displaying SOC is their self-similarity
that derives to a power-law distribution over avalanche sizes.
Respectively, SOC models are mostly studied by
making use of the scaling-type arguments
supplemented with extensive computer simulations (see \cite{2}).
On the contrary, in this paper we put forward an analytical approach,
which is enable to describe in phenomenological manner
both process of a single avalanche formation
and behavior of whole avalanche ensemble.

The SOC behavior appears in vast variety of systems
such as a real sand pile
(ensemble of grains of sand moving
along increasingly tilted surface) \cite{2a} -- \cite{2cc},
intermittency in biological evolution \cite{3},
earthquakes and forest-fires, depinning transitions
in random medium and so on (see \cite{4}).
Among the above models the sandpile is the simplest and most widely
studied as analytically \cite{5}, \cite{6}, so numerically
\cite{1} -- \cite{8}.
In analytical direction, a variety of field theory approaches
is worthwhile to notice.
Among them, the field scheme \cite{9}
based on nonlinear diffusion equation has
failed to account for the main feature of self-organizing
systems -- the self-consistent character of avalanche dynamics.
The visible reason is that using one-parameter approach
does not take into account a feedback between open subsystem
and environment that are related
to order and control parameters, respectively
(see criticism in Refs. \cite{2b}, \cite{2c} also).
Much more substantial picture is given within
two-parameter approaches \cite{2b} -- \cite{2cc}, \cite{10}
that use both fundamental fields: gauge ones related
to hydrodynamical modes type of sand pile height and material
fields as a number of moving sand grains (avalanche size).
Making use of the mean-field approximation shows
that the self-similar regime of the sand pile dynamics
is relevant to subcritical behavior, where a characteristic time
of the order parameter variation is much more than the same
for the control parameter, and the latter follows
the former adiabatically.
Such type adiabatic behavior is inherent in usual regime of system evolving
in course of phase transitions \cite{12} and jammed motion of vehicles \cite{J}, so that
adiabatic approach will be put on basis of our consideration.

Perfect treatment of the SOC has been achieved
within three-parameter approach
based on the Reggeon field theory that
uses the density of active sites $\rho_a$ as the
order parameter and conserved field
of the energy density $\zeta$ as the control parameter \cite{11}, \cite{11a}.
Along this line, SOC regime appears as a result of
competition between
a rate of the energy input $h > 0$ and a dissipation rate $\epsilon$.
It is appeared that the system under consideration
behaves in quite different manner
in case of fixed energy, when $h=\epsilon=0$ and total energy is conserved, and
for driven sandpile, when $h\to 0^{+}$, $\epsilon\to 0$ at stationarity condition $\epsilon > h$.
The first case shows \cite{11}, \cite{11a} to be
reduced to the picture of supercritical regime,
where nonhomogeneity of initial distribution of energy arrives at
a non-Markovian term and space-dependent parameters of theory.
At dimensions above critical value $d_c = 4$,
this case is found to be identical to the simplest
Landau picture
with $\rho_a\sim\zeta -\zeta_c$ in active stationary state
$(\zeta >\zeta_c)$ and $\rho_a = 0$ in absorbing configuration $(\zeta <\zeta_c)$.
Fundamentally different picture appears in the case of driven sandpile,
where due to external input $h\to 0^{+}$ the energy density
is lost as an independent field being reduced to the critical value $\zeta_c$.
Here, the average magnitude of the density of active sites is equal
$\langle\rho_a\rangle = h/\epsilon$,
so that the susceptibility $\chi\equiv\langle\partial\rho_a/\partial h\rangle$ is turned out
to be $\chi =\epsilon^{-1}$.
Respectively, a response function behaves at large distances $r$ as
$\chi(r)\sim r^{2-d}{\rm e}^{-r/\xi}$, where $d$ is a space dimension,
$\xi\sim\epsilon^{-1/2}$ is a correlation length being scale of a system size $L\sim\epsilon^{-1/\mu}$.
It is remarkable that such mean-field-type behavior is caused
solely by stationary condition and translational invariance
\cite{11a}.  A set of basic critical exponents governing scaling
avalanche formation reads \cite{11}:  $\beta=\gamma=\delta=1$,
$\mu=2$, $\nu=1/2$ and $\eta=0$.  On the other hand, scaling
relations accompanied by equality of the susceptibility and mean size
of avalanche yield the following expressions
\begin{equation}
\tau=1+\frac{z}{D},\quad
\tau=2\left(1-\frac{1}{D}\right);\qquad
D=\frac{\mu}{\sigma}
\label{0a}
\end{equation}
for the exponents of the avalanche size distribution
\begin{equation}
P(s, \epsilon)=s^{-\tau}{\cal P}(x);\qquad x\equiv s/s_c, \quad s_c\sim \epsilon^{-1/\sigma},
\label{0b}
\end{equation}
where a critical size $s_c$ is connected with system size  $L\sim\xi$
and a characteristic time $t_c\sim L^{z}$ as $s_c\sim L^{D}\sim t_c^{D/z}$
(exponents $D=\mu/\sigma$ and $z$ are fractal dimension and dynamical exponent
related to a critical avalanche, respectively).
According to Ref. \cite{11}, the mean-field magnitudes of the above exponents are as follows:
$\tau=3/2$, $\sigma=1/2$, $D=4$ and $z=2$.

Along the standard approach \cite{11b}, we will use
as gauge, so material fields:
being reduced to velocity components and sand pile slope, the formers are considered
at study of a single avalanche formation,
whereas the latters are reduced to the number of moving sand grains
at examination of a distribution over avalanche sizes.
Section\ \ref{sec:level2} contains the self-consistent theory of the flux steady-state
developed along the first direction. It enables us to treat
the problem of a single avalanche formation on the basis of the unified
analytical approach
being relevant to the case of fixed energy in Refs. \cite{11a}.
Section\ \ref{sec:level3} deals with accounting
additive noises of the sand velocity components
and sand pile slope.
The noise intensities increase is shown to make possible
avalanche emergence even in non-driven systems.
By this, the control parameter noise plays a crucial role.
Such type fluctuational regime corresponds to the case $h\to 0^{+}$ \cite{11a}, where
a distribution of order parameter is appeared
in an algebraic form with integer exponent.
In order to be not restricted by such particular case,
we introduce unified Lorenz system with fractional feedback in Section\ \ref{sec:level4}.
Using this supposition allows us to describe subcritical regime of the avalanche formation
in natural manner.
The above generalization
is put forward basis of Section\ \ref{sec:level5} devoted to consideration of avalanche ensemble.
Following famous Edwards paradigm \cite{11c}, \cite{11d}, an effective scheme
addressed to nonextensive thermodynamics \cite{Ts} is proposed
to be determined by a time-dependent distribution over energies of moving sand grains.
To generalize the Edwards scheme to nonstationary  nonextensive system,
we use the fractional Lorenz system, where the avalanche size plays a role
of the order parameter, nonextensive complexity is reduced to the conjugate field
and nonconserved energy of the moving grains is
the control parameter. Within the framework of such approach,
the phase diagram is built to define the different
domains of system behavior as a function of noise intensities of above values.
As a result, we arrive at natural conclusion that the power-law
distribution (\ref{0b})
inherent in the SOC regime is caused by noise of the energy.

Section\ \ref{sec:level6} shows that
this distribution is the solution of both nonlinear Fokker-Planck equation,
which appears at studying nonextensive systems \cite{Ts},
and fractional Fokker-Planck equation inherent in L\'evy-type processes
characterized by dynamical exponent $z$ \cite{16}.
As a result, we obtain new relations between the exponent $\tau$ of
the distribution (\ref{0b}), fractal dimension $D$ of phase
space, characteristic exponent of multiplicative noise, number of
governing equations being needed to present self-consistent
behavior in SOC regime, dynamical exponent $z$ and Tsallis
nonextensivity parameter $q$.

Section\ \ref{sec:level7} is Appendix containing basic properties of
fractional integral and derivative, as well as Jackson derivative.

\section{Noiseless avalanche formation}\label{sec:level2}

Within framework of the simplest model of a real sand pile, a dependence $y=y(t,x)$
defines its surface at given instant of time $t$. Locally the flow of
sand can be described in terms of three quantities: the horizontal
and vertical components of the sand velocity,
$\dot{x}=\partial x/ \partial t $,
$\dot{y}=\partial y/ \partial t $, and the surface slope
$y^{\prime}=\partial y/ \partial x $. The key point of our approach
is that the above degrees of freedom are assumed to be of dissipative
type, so that, when they are not coupled, their relaxation to the
steady state is governed by the Debye-type equations:
\begin{equation}
\frac{{\rm d}\dot{x}}{{\rm d}t}=-\frac{\dot{x}}{\tau_{x}},
\label{1a}
\end{equation}
\begin{equation}
\frac{{\rm d}\dot{y}}{{\rm d}t}=-\frac{\dot{y}}{\tau_{y}^{(0)}},
\label{1b}
\end{equation}
\begin{equation}
\frac{{\rm d}y^{\prime}}{{\rm d}t}=
\frac{y^{\prime}_{0}-y^{\prime}}{\tau_{S}},
\label{1c}
\end{equation}
where $\tau_{x},\,\tau_{y}^{(0)}$ and $\tau_{S}$ are the relaxation
times of the velocity components and the slope, respectively.
Eqs.~(\ref{1a}) -- (\ref{1c})
imply that the sand is at rest in the stationary state,
$\dot{x}=\dot{y}=0$ and the equilibrium slope
$y^{\prime}=y^{\prime}_{0}\neq 0$ plays the role of a control
parameter.

Since the motion of sand grain along different directions is not
independent, Eq.~(\ref{1a}) should be changed by adding the term
$f=\dot{y}/ \gamma$ due to liquid friction force along the
$y$-axis ($\gamma$ being the kinetic coefficient). Then, we have

\begin{equation}
\tau_{x} \ddot{x}= -\dot{x}+a^{-1}\dot{y},
\label{2}
\end{equation}
where $a\equiv \gamma/\tau_{x}$.
Note that, owing to the diffusion equation
$\dot{y}=Dy^{\prime\prime}$ ($D$ is the diffusion coefficient), the
friction force appears to be driven by the curvature of the
sand pile surface:

\begin{equation}
f=(D/ \gamma)y^{\prime\prime}.
\label{3}
\end{equation}
On the other hand, when $\ddot{x}=0$ (stationary state), solution
of Eq.~(\ref{2}) defines the tangent line $y=ax+{\rm const}$, so that the
friction force  $f=\tau^{-1}_{x}\dot{x}$
is proportional to the horizontal component
of the sand velocity.
Taking into consideration
the relation (\ref{3}) and using the chain rule
${\rm d}y^{\prime}/{\rm d}t=
\dot{y}^{\prime}+y^{\prime\prime}\dot{x}$,
from Eq.~(\ref{1c})
one obtains the equation of motion for the slope:

\begin{equation}
\tau_{S}\dot{y}^{\prime}=(y^{\prime}_{0}-y^{\prime})-
\left(\tau_{S}/D \right)\dot{y}\dot{x}.
\label{4}
\end{equation}
Following the same line, the equation for the vertical
component of the velocity can be deduced

\begin{equation}
\tau_{y}\ddot{y}=-\dot{y}+
\frac{\tau_{y}}{\tau_{x}}y^{\prime}\dot{x},\qquad
\frac{1}{\tau_{y}}\equiv\frac{1}{\tau^{(0)}_{y}}
\left(
1+\frac{y^{\prime}_{0}}{a}\frac{\tau^{(0)}_{y}}{\tau_{x}}
\right).
\label{5}
\end{equation}
Note the higher order terms are disregarded in Eq.~(\ref{5}) and the
renormalized relaxation time $\tau_{y}$ depending on the
stationary slope $y^{\prime}_{0}$ is introduced.

Equations (\ref{2}), (\ref{4}), (\ref{5}) obtained constitute the basis for
self-consistent description of the sand flow on the surface with the
slope $y^{\prime}$ driven by the control parameter $y^{\prime}_{0}$.
The distinguishing feature of these equations is that nonlinear terms,
that enter Eqs.~(\ref{4}), (\ref{5}), are of opposite signs, while
Eq.~(\ref{2}) is linear. Physically, the latter means that on the
early stage the avalanche begins motion along the tangent
$y=ax+{\rm const}$. The negative sign of the last term in Eq.~(\ref{4}) can
be regarded as a manifestation of Le Chatelier principle, i.e., since
the slope increase results in the formation of an avalanche,
the velocity components $\dot{x}$ and $\dot{y}$ tend to impede  the
growth of the slope. The positive feedback of $\dot{x}$ and
$y^{\prime}$ on $\dot{y}$ in Eq.~(\ref{5})
plays a fundamental part in the problem. As we shall see later, it is
precisely the reason behind the self-organization that brings about
the avalanche generation.

After suitable rescaling, Eqs.~(\ref{2}), (\ref{4}), (\ref{5}) can be
rewritten in the form of the well-known Lorenz system:
\begin{eqnarray}
\dot{u}&=&-u+v,\label{6a}\\
\epsilon\dot{v}&=&-v+uS,\label{6b}\\
\delta\dot{S}&=&(S_{0}-S)-uv, \label{6c}
\end{eqnarray}
where $u{\equiv} (\tau_{y}/\tau_{x})^{1/2}(\tau_{S} /D)^{1/2}\dot{x}$,
$v{\equiv} (\tau_{y}/\tau_{x})^{1/2}(\tau_{S}
/D)^{1/2}\dot{y}/a$ and $S{\equiv}(\tau_{y}/
\tau_{x})y^{\prime}/a$ are the dimensionless velocity
components and the slope, respectively;
$\epsilon\equiv\tau_{y}/
\tau_{x},\, \delta\equiv\tau_{S}/ \tau_{x}$  and the  dot now stands for
the derivative with respect to the dimensionless time $t/\tau_{x}$.
In the general case, the system
(\ref{6a})--(\ref{6c})
cannot be solved analytically, but
in the simplest case, where $\epsilon\ll 1\, {\rm and}\, \delta\ll
1$, the vertical velocity $v$ and the slope $S$ can be eliminated by
making use the adiabatic approximation that implies neglecting of
the left hand side of Eqs.~(\ref{6b}), (\ref{6c}). As a result, the
dependencies of $S$ and $v$ on the horizontal velocity $u$ are given
by
\begin{equation}
S=\frac{S_{0}}{1+u^{2}},\qquad v=\frac{S_{0}u}{1+u^{2}}.
\label{7}
\end{equation}
Note that, under $u$ is within the physically meaningful range between $0$
and $1$, the slope $S$ is a monotonically decreasing function of $u$,
whereas the velocity $v$ increases with $u$
(at $u>1$ we have ${\rm d}v/{\rm d}u<0$ and the flow of the sand
becomes unstable).

Substitution of the second Eq.~(\ref{7}) into Eq.~(\ref{6a}) yields the
Landau-Khalatnikov equation
\begin{equation}
\dot{u}=-\frac{\partial E}{\partial u}
\label{8}
\end{equation}
with the kinetic energy
\begin{equation}
E=\frac{1}{2}u^{2}-
\frac{1}{2}S_{0}\ln{\left(1+u^{2}\right)}.
\label{9}
\end{equation}
For $S_{0}<1$, the $u$-dependence of $E$ is monotonically
increasing and the only stationary value of $u$ equals zero,
$u_{0}=0$, so that there are no avalanches in this case.
Obviously, such steady-state relevant to absorbing configuration
studied in Ref. \cite{11a}.
If the slope
$S_{0}$
exceeds the critical value, $S_{c}=1$, the kinetic energy assumes the
minimum with nonzero steady state
velocity components $u_{e}=v_{e}=(S_{0}-1)^{1/2}$ and
the slope $S_{e}=1$.

The above scenario represents supercritical regime of the avalanche
formation and addresses to the second-order phase transition  \cite{10}. The
latter can be easily seen from the expansion of the kinetic energy
(\ref{9}) in power series of $u^{2}\ll 1$. So the critical
exponents $\gamma$, $\delta$, $\nu$ are identical to those obtained within the framework of
the mean-field theory  \cite{11}.
However, the magnitude $\beta=1/2$ is twice as little because
our order parameter (being the velocity) is not reduced to the same (the number of active sites)
in theory \cite{11}.

The drawback of the outlined approach is that it fails to account for
the subcritical regime of the self-organization that is the reason
for the appearance of avalanches  and analogous to the
first-order phase transition, rather than the second-order one.
So one has to modify the above theory by assuming that
the effective relaxation time $\tau_{x}(u)$
increases with velocity $u$ from value $\tau_{x}(1+m)^{-1}$,
$m>0$ to  $\tau_{x}$ \cite{12}.
The simplest two-parameter approximation  is
\begin{equation}
\frac{\tau_{x}}{\tau_{x}(u)}=1+\frac{m}{1+(u/u_{0})^{2}},
\label{10}
\end{equation}
where $0<u_{0}<1$. The expression for the
kinetic energy (\ref{9}) then changes by adding the term
\begin{equation}
\Delta E=\frac{m}{2}u_{0}^{2}
\ln{\left[1+\left(\frac{u}{u_{0}}\right)^{2}\right]}
\label{11}
\end{equation}
and the stationary values of $u$ are following:
\begin{eqnarray}
u_{e}^{m}&=&u_{00}
\left\{
1\mp \left[
1+u_{0}^{2}u_{00}^{-4}(S_{0}-S_{c})
\right]^{1/2}
\right\}^{1/2}, \label{12}\\
2u_{00}^{2}&\equiv& (S_{0}-1)+S_{c}u_{0}^{2},\qquad
S_{c}\equiv1+m.\nonumber
\end{eqnarray}
The upper sign in the right hand side of Eq.~(\ref{12}) meets the
value of the unstable state $u^{m}$, where the kinetic energy
$E+\Delta E$ has the maximum, the lower one corresponds to the stable
state $u_{e}$. The corresponding value of the stationary slope
\begin{equation}
S^{m}=\frac
{1+u_{00}^{2}+\sqrt
{\left(1+u_{00}^{2}\right)^{2}-
\left(1-u_{0}^{2}\right)S_{0}}}
{1-u_{0}^{2}}
\label{13}
\end{equation}
smoothly increases from the value

\begin{equation}
S_{{\rm min}}=1+u_{0}\sqrt{m/(1-u_{0}^{2})}
\label{14}
\end{equation}
at the parameter $S_{0}=S_{c0}$ with
\begin{equation}
S_{c0}=\left(1-u_{0}^{2}\right)S_{{\rm min}}^{2}
\label{15}
\end{equation}
to the marginal value $S_{c}=1+m$ at $S_{0}=S_{c}$. The
$S_{0}$-dependencies of $u_{e},\,u_{m},\,{\rm and}\,S_{e}$
are presented in
Fig.~1. As is shown in Fig.~1a, under the adiabatic condition
$\tau_{S}\ll\tau_{x}$ is met and the parameter $S_{0}$ slowly
increases being below $S_{c}$ ($S_{0}\le S_{c}$), no avalanches can
form. At the point $S_{0}=S_{c}$ the velocity $u_{e}$ jumps upward
to the value $\sqrt{2}u_{00}$ and its further smooth
increase is determined
by Eq.~(\ref{12}). If the parameter $S_{0}$ then goes downward the
velocity $u_{e}$ continuously decreases to the point, where
$S_{0}=S_{c0}\,{\rm and}\,u_{e}=u_{00}$. At this point the velocity
jump-like goes down to zero. Referring to Fig.~1b, the stationary
slope $S_e$ shows a linear increase from $0$ to $S_{c}$ with the parameter
$S_{0}$ being in the same interval and, after the jump down to
the value $(1-u_{0}^{2})^{-1}$ at $S_{0}=S_{c}$, $S_{e}$ smoothly
decays to $1$ at
$S_{0}\gg S_{c}$. Under the parameter $S_{0}$ then decreases from
$S_{c}$ down to $S_{c0}$, the slope grows.
When the point (\ref{15}) is reached, the avalanche
stops, so that the slope undergoes the jump from the value
(\ref{14}) up to the one defined by Eq.~(\ref{15}). For
$S_{0}<S_{c0}$ again the parameter
$S_{e}$ does not differ from $S_{0}$.
Note that this subcritical regime is realized provided the parameter
$m$, that enters the dispersion law (\ref{10}), is greater than
\begin{equation}
m_{{\rm min}} = \frac{u_{0}^{2}}{1-u_{0}^{2}}.
\label{16}
\end{equation}

Clearly, according to the picture described, the avalanche generation
is characterized by the well pronounced hysteresis, when the grains
of sand initially being at rest begin to move downhill only if the
slope of the surface exceeds its limiting value $S_{c}=1+m$, whereas
the slope $S_{c0}$ needed to stop the avalanche is less
than $S_{c}$ (see Eqs.~(\ref{14}), (\ref{15})). This is the case in the
limit $\tau_{S}/\tau_{x}\to 0$ and the hysteresis loop shrinks with
the growth of the adiabaticity parameter
$\delta\equiv\tau_{S}/\tau_{x}$.  In addition to the smallness of
$\delta$, the adiabatic approximation implies the ratio
$\tau_{y}/\tau_{x}\equiv\epsilon$ is also small. In contrast to the
former, the latter does not seem to be realistic for the system under
consideration, where, in general, $\tau_{y}\approx\tau_{x}$. Thus, it is
of interest to study to what extent the finite value of $\epsilon$
could change the results.

Owing to the condition $\delta\ll 1$, Eq.~(\ref{6c}) is still algebraic and
$S$ can be expressed in terms of $u$ and $v$. As a result, we derive
the system of two nonlinear differential equations that can be
studied by the phase portrait method \cite{12}. The phase portraits
for various values of $\epsilon$ are displayed in Fig.~2, where the
node point $O$ represents the stationary state and the saddle
point $S$ is related to
the maximum of the kinetic energy. As is seen from the figure,
independently of $\epsilon$, there is the universal
section that
attracts all phase trajectories and its structure is appeared to be
almost insensitive to changes in $\epsilon$. Analysis of time
dependencies $v(t)$ and $u(t)$ reveals that the velocity components
slow down appreciably on this section in comparison to the rest parts
of trajectories that are almost rectilinear (it is not difficult to
see that this effect is caused by the smallness of the parameter $\delta$). Since
the most of time the system is in vicinity of this universal section,
we arrive at the conclusion that finite values of $\epsilon$ do not
affect qualitatively the above results obtained in the adiabatic
approximation.

\section{Noise influence on avalanche formation}\label{sec:level3}

We focus now on studying the affect of
additive noises of the velocity components
$u$, $v$, and the slope $S$. With this aim, we should add
to right hand parts of Eqs.~(\ref{6a}) -- (\ref{6c}) the stochastic terms
$I_u^{1/2}\xi$, $I_v^{1/2}\xi$, $I_S^{1/2}\xi$, respectively
[here the noise intensities $I_{u}$, $I_{v}$, $I_{S}$ are measured in units
$(\tau_{x}/\tau_{y})(D/\tau_{S})$,
$a^2(\tau_{x}/\tau_{y})(D/\tau_{S})$,
$a^2(\tau_{x}/\tau_{y})$, correspondingly, and
$\xi(t)$ is $\delta$-correlated stochastic function] \cite{13}.
Then, within the adiabatic approximation,
equations (\ref{6b}), (\ref{6c}) could be reduced to the
time-dependencies
\begin{eqnarray} &v&(t)=\bar v + \tilde
v\xi(t),\quad S(t)=\bar S + \widetilde S\xi(t); \label{X1}\\
\nonumber\\
&\bar v &\equiv S_0 ud(u),\quad
\tilde{v}\equiv\sqrt{I_v+I_S u^2}~d(u),\label{X2}\\
&\bar S &\equiv S_0 d(u),\quad
\widetilde{S}\equiv\sqrt{I_S + I_v u^2}~d(u),\qquad d(u)\equiv(1+u^2)^{-1}.\nonumber
\end{eqnarray}
Here deterministic components are reduced to Eqs.~(\ref{7}), whereas
fluctuational ones follow from the known property of additivity
of variances of independent Gaussian random quantities \cite{13}.
Thus, using slaving principle inherent in synergetics \cite{H}
transforms noises of both vertical velocity component $v$ and slope
$S$, having be adiabatic initially, to multiplicative form.  As a
result, combination of Eqs.~(\ref{6a}), (\ref{X1}), (\ref{X2})
arrives at Langevin equation \begin{equation} \dot u = f(u) +
\sqrt{I(u)}~\xi(t),\quad f\equiv-~{\partial E\over\partial u},
\label{VI.5_17a}
\end{equation}
where force $f$ is related to the energy $E$ determined by Eq.~(\ref{9}) and
expression for effective noise intensity
\begin{equation}
I(u)\equiv I_u + \left(I_v + I_S~u^2\right)d^2(u)
\label{X3}
\end{equation}
is obtained in accordance with above mentioned property of noise variances additivity.
In order to avoid mistakes, one should notice that direct insertion of Eqs.~(\ref{X1}), (\ref{X2}) into (\ref{6a})
results in appearance of stochastic addition
\begin{equation}
\left[I_u^{1/2} + \left(I_v^{1/2} + I_S^{1/2}u \right)d(u)
\right]\xi(t),
\label{X4}
\end{equation}
whose squared amplitude is quite different from
effective noise intensity (\ref{X3}).
Moreover, contrary to expressions (\ref{X2}), direct using adiabatic approximation
in Eqs.~(\ref{6b}), (\ref{6c}) reduces fluctuational additions in Eqs.~(\ref{X1}) to the forms:
$\tilde{v}\equiv(I_v^{1/2}+I_S^{1/2} u)d(u)$,
$\widetilde S\equiv(I_S^{1/2} - I_v^{1/2} u)d(u)$.
The latter is obviously erroneous since the
effective noise of the slope $\widetilde S$ disappears entirely
for the horizontal velocity $u=\sqrt{I_S/I_v}$.
The reason of such contradiction is that Langevin equation, being stochastic in nature,
does not permit usage of usual analysis methods (see \cite{13}).

To continue along standard direction, let us write the Fokker-Planck equation
related to Langevin equation (\ref{VI.5_17a}):
\begin{equation}
{\partial P(u, t)\over\partial t}=
{\partial\over\partial u}\left\{-f(u) P(u, t)+{\partial \over\partial u}\left[I(u)P(u, t)\right]\right\}.
\label{X5}
\end{equation}
At steady state, which will be considered only, the probability distribution $P(u, t)$ becomes
time-independent function $P(u)$ and usual condition, that
expression in braces of right hand part in Eq.~(\ref{X5}) is zero,
arrives at stationary distribution
\begin{equation}
P(u)=Z^{-1}\exp\lbrace-U(u)\rbrace,
\label{a}
\end{equation}
where $Z$ is a normalization constant and effective
energy
\begin{equation}
U(u)=\ln I(u)-\int\limits^u_0{f(u')\over I(u')}{\rm d}u',\quad
f\equiv -~{\partial E\over\partial u},
\label{VI.5_17}
\end{equation}
is determined by the bare energy $E$, Eq.~(\ref{9})
and the noise intensity $I(u)$, Eq.~(\ref{X3}) \cite{14}.
Combining these expressions, we might find the explicit form of $U(u)$,
which is too cumbersome to be reproduced here.
The equation which defines the locations of the maxima of
distribution function $P(u)$
\begin{equation}
x^3 - S_0 x^2 - 2I_S x + 4(I_S - I_v) = 0,
\quad x \equiv 1 + u^2, \label{VI.5_19}
\end{equation}
is much simpler.
According to Eq.~(\ref{VI.5_19}), they are insensitive to changes
in the intensity of noise $I_u$ of the velocity component $u$,
but are determined by the value $S_0$
of the sand pile slope and the intensities
$I_v$, $I_S$ of the noises of vertical velocity component $v$
and slope $S$, which acquire the multiplicative character
in Eq.~(\ref{X3}).
Hence, it can be put for simplicity $I_u=0$
and Eqs.~(\ref{9}), (\ref{VI.5_17}), (\ref{X3}) give
the following expression for the effective energy:
\begin{eqnarray}
U(u)&=&{1\over 2}\left[{u^4\over 2} +
(2-S_0-i)u^2+\right. \label{VI.5_27}\\
&&\left. (1-i)\left(1-S_0-i
\right)\ln(i+u^2)\right]+
I_S\ln[g_S^2(u)+i g_v^2(u)], \quad
i\equiv{I_v/I_S}.
\nonumber
\end{eqnarray}

According to Eq.~(\ref{VI.5_19}),
the effective energy (\ref{VI.5_27}) has a minimum at $u=0$
if the driven slope $S_0$
does not exceed the critical level
\begin{equation}
S_c=1+2 I_S-4I_v,\label{VI.5_29}
\end{equation}
whose value increases at
increasing intensity of noise of the sand pile slope, but
decreases with one of the vertical velocity.
Here, sand grains are at rest.
In the simple case $I_v=0$, the avalanche creation
is related to solutions
\begin{equation}
u^2_\pm={1\over
2}\left[S_0-3+\sqrt{(3-S_0)^2+4(2S_0-3+2 I_S)}\right],
\label{VI.5_34}
\end{equation}
which are obtained from Eq.~(\ref{VI.5_19}) after
elimination of the root $u^2=0$.
The magnitude of this solution has its minimum
\begin{equation}
u_c^2={1\over 2}\left[
(S_0-3)-\sqrt{(S_0 +7)(S_0-1 )}\right]
\label{VI.5_35}
\end{equation} on the line defined by expression
(\ref{VI.5_29}) with $I_v=0$. At $S_0 < 4/3$
the roots $\pm u_c$ are complex,
at $S_0 =4/3$ they become zero, at $S_0 >4/3$ they are real
and  $ u_+=- u_-$.
In this way, the tricritical point
\begin{equation} S_0
= 4/3,\qquad  I_S = 1/6 \label{VI.5_36}
\end{equation}
addresses to the appearance of roots
$ u_\pm\ne 0$ of Eq.~(\ref{VI.5_19})
(avalanche creation).
If condition (\ref{VI.5_29}) is satisfied,
the root $ u=0$ corresponds to the minimum of the
effective energy (\ref{VI.5_27})
at $S_0 < 4/3$, whereas at $S_0 > 4/3$
this root corresponds to the maximum, and the roots   $ u_\pm$ -- to
symmetrical minima.

Now, we find another condition of stability of roots
$u_\pm$.  Setting the discriminant of Eq.~(\ref{VI.5_19}) equal to
zero, we get the equations
\begin{equation}
I_S=0,\qquad I_S^2-I_S
\left[{27\over 2}\left(1-{S_0\over 3}\right)-{S_0^2\over 8}\right]+{S_0^3\over 2}=0,
\label{VI.5_37}
\end{equation}
the second of which gives
\begin{equation}
2I_S=\left[{27\over 2}\left(1-{S_0\over 3}\right) -{S_0^2\over
8}\right] \pm  \left\{ \left[ {27\over 2}\left(1-{S_0\over
3}\right) -{S_0^2\over 8}\right]^2-
2S_0^3\right\}^{1/2}.\label{VI.5_38}
\end{equation}
This equation
defines a bell-shaped curve $S_0(I_S)$, which intersects
the horizontal axis at the points $I_S=0$  and
$I_S=27/2$, and has a maximum at
\begin{equation}
S_0=2, \qquad I_S=2.  \label{VI.5_39}
\end{equation}
It is easy to see that for $I_v=0$
this line touches the curve (\ref{VI.5_29}) at point
(\ref{VI.5_36}).

Let us consider now the more general case of two
multiplicative noises $I_v, I_S\ne 0$.
Introducing the parameter $a=1-i$, $i\equiv I_v/I_S$ and the renormalized
variables $\tilde{I}\equiv {I_S/ a^2}$,
$\tilde{S_0}\equiv{S_0/ a}$, $ \tilde{u}^2 =
(1+ u^2)/a-1$, at $i<1$ we
reproduce all above expressions
with the generalized energy $\widetilde{U}/\tilde{I}$
in Eq.~(\ref{VI.5_27}).
Thus, action of the noise of the vertical velocity component
$v$ is reduced to the
renormalization of the extremum value of the horizontal one
by the quantity $(a^{-1}-1)^{1/2}$. As a result, the region of divergence
$\tilde{u}\approx 0$ becomes
inaccessible.

The condition of extremum of the generalized energy (\ref{VI.5_27}) splits
into two equations, one of which is simply  $u=0$, and the other
is given by Eq.~(\ref{VI.5_19}). As mentioned above,
analysis of the latter indicates
that the line of existence of the zero solution is defined by
expression (\ref{VI.5_29}). The tricritical point  has the coordinates
\begin{equation} S_0 ={4\over 3}(
1- I_v),\quad  I_S ={1\over 6}\left( 1
+8 I_v\right). \label{VI.5_48}
\end{equation}
The phase diagram for the fixed intensities $I_v$ is shown in Fig.~3.
Here the curves $1$, $2$ define the thresholds of absolute loss of stability
for the fluxless and flux steady-states, respectively.
Above line $1$ the system occurs in a stable flux state,
below curve $2$ it is in fluxless one, and between these lines
the two-phase domain is realized.
For $I_v<1/4$ situation is generally the same as
in the simple case $I_v=0$ (see Fig.~3a).
At $I_v>1/4$ the avalanches formation is possible even
for small intensities $I_S$ of the slope noise (Fig.~3b).
According to (\ref{VI.5_48}), the tricritical point
occurs on the $I_S$-axis at $I_v=1$, and for the noise intensity $I_v$
larger than the critical value $I_v=2$
the stable fluxless state disappears (see Fig.~3c).

The above consideration shows that the dissipative dynamic
of grains flow in a real sand pile can be represented
within the framework of the Lorenz model,
where the horizontal and vertical velocity components
play roles of the order parameter and its conjugate field, respectively,
and the sand pile slope is the control parameter.
In Section\ \ref{sec:level2}, the noiseless case is examined to show that an avalanche
is created
if the externally driven sand pile slope $y'_0$ is larger than the
critical magnitude
\begin{equation}
y'_c=(\tau_x\tau_y)^{-1/2}\gamma.
\label{ba}
\end{equation}
In this sense, the systems
with small values of the kinetic coefficient $\gamma$
and large relaxation times $\tau_x$, $\tau_y$
of the velocity components are preferred.
However, the sand flow appears here as a phase transition
because the avalanche creation
in the noiseless case is possible only due to the externally driven
growth of the sand pile slope.

Accounting the additive noises of the above degrees of freedom
shows that the stochasticity influence
is non-essential for the horizontal velocity component
and is crucial for both the vertical one and the sand pile slope.
The avalanche appears spontaneously if the dimensionless noise intensities
are connected by linear relation
\begin{equation}
I_{S}=- {1\over 2} + 2I_{v},
\label{bbb}
\end{equation}
following from Eq.~(\ref{VI.5_19}) at conditions $x=1$ $(u=0)$, $S_0=0$.
According to Eq.~(\ref{bbb}), in absence of the sand pile slope noise
the avalanche is created if the intensity of
vertical velocity component exceeds the value
\begin{equation} I_{v0}={1\over
4}~{D\gamma^2\over\tau_x\tau_y\tau_S},
\label{b}
\end{equation}
corresponding to the point $O$ in Fig.~4.
Increase of both the vertical velocity and the sand pile slope noises
causes avalanche formation if its intensities are bounded
by the condition (\ref{bbb}).
With further increase of these intensities above magnitudes
\begin{equation}
I_{v1}={D\gamma^2\over\tau_x\tau_y\tau_S},\quad
I_{S1}={3\over 2}{\gamma^2\over\tau_x\tau_y},
\label{bb}
\end{equation}
the domain of the mixed state appears at the point $T$ in Fig.~4.
If the noise intensity of the vertical velocity is
more than the larger value
\begin{equation}
I_{v2}=2~{D\gamma^2\over\tau_x\tau_y\tau_S},
\label{c}
\end{equation}
corresponding to the sand pile slope noise
$I_{S2}=6\gamma^2/\tau_x\tau_y$ (the point $C$ in Fig.~4),
the fluxless steady-state disappears at all.

Physically, we have to take into consideration that
the SOC regime is not relevant to flux-type avalanche state
itself, but rather to an intermittent regime of avalanche formation
corresponding to the domains on the phase diagrams
in Figs.~3, 4, where a mixture of both phases A and N
(avalanche and non-avalanche) exists.
According to above analysis, such an intermittent behavior may be realized
within the region located upper the line (\ref{bbb}) and outside the curve
determined by the equation
\begin{equation}
I_v=I_S\left[1-\left({2\over 27}\right)^{1/2}\sqrt{I_S}\right]
\label{ca}
\end{equation}
for the dimensionless values $I_{v}$, $I_{S}$.
Corresponding phase diagram is depicted in Fig.~4
to show very non-trivial form (especially, within the domain
$I_{v1}\leq I_{v}\leq I_{v2}$).

\section{Generalizing self-similarity}\label{sec:level4}

To proceed the consideration of the system behavior,
let us examine explicit form of the probability (\ref{a}) determined for different regimes
by the effective energy (\ref{VI.5_17}).
In the case $I_u, I_S\ll I_{v}$, we obtain Gibbs-type distribution
\begin{equation}
P(u)\approx I_v^{-1}(1+u^2)^2 \exp\left\{I_v^{-1}\int f(u)
(1+u^2)^2{\rm d}u \right\},\
f(u)\equiv -u+S_0 u/(1+u^2),
\label{d}
\end{equation}
being opposite to the power dependence inherent in self-similar systems.
Contrary, at intermittent behavior, when $I_u, I_{v}\ll I_{S}$,
supercritical values of the slope noise intensity $I_S$ cause
the following distribution form:
\begin{equation}
P(u)\approx I_S^{-1}\left({1+u^2\over u}\right)^2 \exp\left\{I_S^{-1}\int
{f(u)(1+u^2)^2\over u^2}{\rm d}u\right\}
\sim  u^{-2}.
\label{e}
\end{equation}
Thus, the case $I_u, I_{v}\ll I_{S}$ addresses to
the power-law distribution being relevant to self-similar behavior.
However, as corresponding consideration \cite{14} shows,
obtained exponent is not reduced to integer 2, generally.

To get rid off such a restriction, the multiplier $u$ in nonlinear terms
of Eqs.~(\ref{6a}) -- (\ref{6c}) is supposed to be replaced by $u^{a }$,
where an exponent $0\leq a\leq 1$.
With accounting the stochastic additions one obtains then
the basic equations in dimensionless form
\begin{eqnarray}
\dot{u}&=&-u+v+\sqrt{I_u}~\xi(t),\nonumber\\
\epsilon\dot{v}&=&-v+u^{a }S+\sqrt{I_v}~\xi(t),\label{6bb}\\
\delta\dot{S}&=&(S_{0}-S)-u^{a }v+\sqrt{I_S}~\xi(t). \nonumber
\end{eqnarray}
Thus, it is appeared that the agreement of the
Lorenz self-organization
scheme with SOC conception related to self-similar
systems is achieved if one assumes that both positive and negative feedbacks
are fractional in the nature.
Within such a supposition, the adiabatic approximation $\epsilon, \delta\ll $1
arrives at the Langevin equation (cf. Eq.~(\ref{VI.5_17a}))
\begin{eqnarray}
\dot{u}&=f_{a }(u)+\sqrt{I_{a }(u)}\xi(t),
\label{9ad}
\end{eqnarray}
where force $f_{a }(u)$ and noise intensity $I_{a }(u)$ are as follows:
\begin{eqnarray}
&f_{a }(u)\equiv -u+S_0 u^{a }d_{a }(u),&\nonumber\\
&I_{a }(u)\equiv I_u + \left(I_v + I_S u^{2a}\right)d^2_{a }(u),\quad
d_{a }(u)\equiv\left(1+u^{2a }\right)^{-1}.&
\label{9a}
\end{eqnarray}
Corresponding distribution (cf. Eqs.~(\ref{a}), (\ref{VI.5_17}))
\begin{eqnarray}
P_{a }(u)={Z^{-1}\over I_{a }(u)}\exp\left\{-E_a(u)\right\}
\label{9p}
\end{eqnarray}
with partition function $Z$ is determined by effective potential
\begin{eqnarray}
E_a(u)\equiv -\int\limits^u_0{f_{a }(u')\over
I_{a }(u')}{\rm d}u'.
\label{9pp}
\end{eqnarray}
Extremum points of this distribution are determined by equation
\begin{eqnarray}
2aI_S u^{2a}+\left(1+u^{2a}\right)^2 u^{1-a}
\left[S_0-u^{1-a}\left(1+u^{2a}\right)\right]
&=&2a(I_S-2I_v),
\label{9x}
\end{eqnarray}
which gives the boundary of the flux state
\begin{eqnarray}
I_S = 2 I_v,
\label{9z}
\end{eqnarray}
related to $u=0$. Critical values of state parameters are fixed
by condition $|{{\rm d}u\over{\rm d}S_0}| =\infty$ to arrive at
additional equation
\begin{eqnarray}
 u^{2(1-a)}\left(1+u^{2a}\right)^2
\left[\left(2+a^{-1}\right)+
\left(a^{-1}-1\right)u^{-2a}\right]&-&\nonumber\\
{1\over 2}S_0 u^{1-a}\left(1+u^{2a}\right)
\left[\left(3+a^{-1}\right)+
\left(a^{-1}-1\right)u^{-2a}\right]
&=&2aI_S.
\label{9y}
\end{eqnarray}
Expressions (\ref{9x}) -- (\ref{9y}) generalize the simple equalities
(\ref{VI.5_19}), (\ref{bbb}) and (\ref{ca}) related to the case $a =1$.

Above expressions show that qualitative results of Section\ \ref{sec:level3}
obtained for the particular case
$a=1$ are kept invariant at passage to
general case $0 \leq a \leq 1$.
Indeed, the most essential difference is
observed for noiseless case,
namely the steady-state velocity $u$ becomes nonzero within
whole interval of the driven slope $S_0$ (see Fig.~5).
Increase of the vertical velocity noise $I_v$ causes monotonic $u$-growth,
whereas $I_S$-increase arrives at an effective barrier formation near
the point $u=0$, so that the dependence $u(S_0)$ becomes non-monotonic
in character at magnitudes $I_S$ above the straight line (\ref{9z}) (see Figs.~6).
Here, by analogy with noiseless case (see Fig. 1),
lower branches of curves correspond to unstable
magnitudes of the order parameter, the uppers -- to stable ones.
According to Fig.~7, the domain, where avalanches can
not be created,
is located near an intermediate magnitudes of the state parameters
$S_0$, $I_v$, $I_S$.
The phase diagram related to the avalanche formation reveals the same form
as for the simplest case $a =1$, but the straight line
(\ref{bbb}) shifts abruptly to (\ref{9z}) with escaping the point $a =1$
(compare Fig.~8 with Fig.~4).
According to Fig.~9, increase of the vertical velocity noise $I_v$
increases the domain of the avalanche
formation.

\section{Size distribution in self-similar ensemble of
avalanches}\label{sec:level5}

Contrary to the previous, when the process of a single avalanche
formation has been considered,
further we aim to study analytically self-similar size
distribution (\ref{0b}) over avalanche ensemble.
This means that, along line of Section\ \ref{sec:level3}, we will account for noises of complete set
of freedom degrees, on the one hand, and the fractional feedback type of introduced in
Section\ \ref{sec:level4},
on the other one. Thereby, the Lorenz system unified in above manner
is the basis of our examination.
However, instead of visible geometric-and-mechanic characteristics
of 'real' sand pile, the system under consideration is parametrized now
by a set of pseudo-thermodynamical values, which describes the avalanche ensemble
in a spirit of the famous Edwards paradigm \cite{11c}, \cite{11d}
generalized to nonstationary system.
In this line, we study time dependencies of the avalanche size,
nonextensive complexity and nonconserved energy of the moving grains.
Within the framework of usual synergetic approach,
these degrees of freedom play roles
of order parameter, conjugate field and control parameter, respectively.

It is very important that using slaving principle of synergetics and
fractional nature of the system feedback
are shown above to stipulate multiplicative character of noise.
It will be appeared below, this causes
a nonextensivity of applied thermodynamical scheme, so that
we have to use $q$-weighted averages instead of usual ones.
So, energy of moving sand grains is defined by expression
\begin{eqnarray}
\zeta_q\equiv\sum_i\zeta_i~ p_i^q,
\label{A}
\end{eqnarray}
where $p_i$ is a probability to move grain $i$ with energy $\zeta_i$,
$q\neq 1$ is a positive parameter being measure of the system nonextensivity
determined below.
Respectively, nonextensive complexity of moving sand grains is an analog
of Tsallis entropy \cite{Ts} to be determined as follows:
\begin{eqnarray}
\Sigma_q\equiv - \frac{\sum_i p_i^q - 1}{q-1}.
\label{B}
\end{eqnarray}
Three-parameter set of standard synergetic scheme \cite{H} is completed
by avalanche size $s$.

Following above elaborated line, we postulate that self-consistent behavior
of the system under consideration
is presented adequately by set of above quantities governed by the Lorenz-type equations
(cf. Eqs.~(\ref{6bb}))
\begin{eqnarray}
\tau_s\dot{s}&=&-s+a_s\Sigma_q+\sqrt{I_s}~\xi(t),\nonumber\\
\tau_{\Sigma}\dot{\Sigma_q}&=&-\Sigma_q+a_{\Sigma}s^{\tau/2}\zeta_q+\sqrt{I_{\Sigma}}~\xi(t),\label{C}\\
\tau_{\zeta}\dot{\zeta_q}&=&(\zeta^{0}-\zeta_q)-a_{\zeta}s^{\tau/2}\Sigma_q+\sqrt{I_{\zeta}}~\xi(t).
\nonumber
\end{eqnarray}
Here $\tau_s$, $\tau_{\Sigma}$, $\tau_{\zeta}$ note relaxation times of corresponding values,
$a_s$, $a_{\Sigma}$, $a_{\zeta}$ are related feedback parameters,
$I_s$, $I_{\Sigma}$, $I_{\zeta}$ are respective noise intensities,
$\tau$ is a positive exponent and
$\zeta^0$ is externally driven energy of sand motion.
The distinguishing feature of the first of these equations is that
in noiseless case genuine characteristics $s$, $\Sigma_q$ are connected in linear manner.
On the other hand, connection of
thermodynamic-type values $\zeta_q$, $\Sigma_q$ with the avalanche size $s$ is stated by
two last equations
(\ref{C}) to be nonlinear in nature. Physically, this means
linear relation between complexity and avalanche size near steady-states.
Running away these derives to
negative feedback of the avalanche size with the complexity that arrives at
the energy decrease -- in accordance with Le Chatelier principle,
and positive feedback between the avalanche size and the energy that brings the complexity increase
to be reason of the avalanche ensemble self-organization.

To analyze the system (\ref{C}), it is convenient to measure the time $t$ in unit $\tau_s$, as well as
the values $s$, $\Sigma_q$, $\zeta_q$ and
$I_s$, $I_{\Sigma}$, $I_{\zeta}$ are related to the scales:
\begin{eqnarray}
&s^{sc}\equiv(a_{\Sigma}a_{\zeta})^{-\frac{1}{\tau}},\
\Sigma_q^{sc}\equiv a_s^{-1}(a_{\Sigma}a_{\zeta})^{-\frac{1}{\tau}},\
\zeta_q^{sc}\equiv a_s^{-1}a_{\Sigma}^{-(\frac{1}{\tau}+\frac{1}{2})}a_{\zeta}^{-(\frac{1}{\tau}-\frac{1}{2})};&
\nonumber\\
&I_s^{sc}\equiv(a_{\Sigma}a_{\zeta})^{-\frac{2}{\tau}},\
I_\Sigma^{sc}\equiv a_s^{-2}(a_{\Sigma}a_{\zeta})^{-\frac{2}{\tau}},\
I_\zeta^{sc}\equiv a_s^{-2}a_{\Sigma}^{-(\frac{2}{\tau}+1)}a_{\zeta}^{-(\frac{2}{\tau}-1)}.&
\label{D}
\end{eqnarray}
Then, the modified Lorenz system (\ref{C}) takes the simple form
\begin{eqnarray}
\dot{s}&=&-s+\Sigma_q+\sqrt{I_s}~\xi(t),\nonumber\\
\vartheta\dot{\Sigma_q}&=&-\Sigma_q+s^{\tau/2}\zeta_q+\sqrt{I_{\Sigma}}~\xi(t),\label{E}\\
\theta\dot{\zeta_q}&=&(\zeta^{0}-\zeta_q)-s^{\tau/2}\Sigma_q+\sqrt{I_{\zeta}}~\xi(t),
\nonumber
\end{eqnarray}
where relaxation time ratios are introduced:
\begin{eqnarray}
\vartheta\equiv\tau_{\Sigma}/\tau_s,\quad
\theta\equiv\tau_{\zeta}/\tau_s.
\label{F}
\end{eqnarray}
It is worthwhile to notice that system (\ref{E}) is passed to the form of Eqs.~(\ref{6bb}) if the values
$s$, $\Sigma_q$, $\zeta_q$, $\tau/2$, $\vartheta$, $\theta$ are replaced by
$u$, $v$, $S$, $a$, $\epsilon$, $\delta$, respectively.

It is well-known that complete set of SOC systems can
be reduced to one of two families \cite{11}:
systems with deterministic dynamics extremely driven by random environment
(growing interface models, Bak-Sneppen evolution model and so one) and
the stochastic dynamics family (models of earthquakes, forest-fire et cetera).
\footnote{Generally, we face here with much more complicated problem,
see Ref.~\cite{15a}.} Remarkable peculiarity of obtained system
(\ref{E}) is a possibility to present in natural manner both mentioned
families. So, the former is related to noiseless case,
when $I_s, I_{\Sigma}, I_{\zeta}=0$ but magnitude of the energy
relaxation time is above than the ones for complexity and
avalanche size ($\tau_\zeta\geq\tau_\Sigma, \tau_s$); on the other
hand, a parameter of environment drive $\zeta^{0}$ has to take
magnitudes above the critical one $\zeta_c=1$ \cite{12}.
In such a case, the system (\ref{E}) describes
strange attractor \cite{H} that may represent
behavior of the first type SOC systems.
Proper stochastic behavior is relevant to nonzeroth noise intensities $I_s, I_{\Sigma}, I_{\zeta}\ne 0$
that make possible the SOC regime appearance even in absence
of driven affect ($\zeta^0=0$).

Taking into account that problem of Lorenz strange attractor is well known \cite{H},
we will restrict ourselves further with treatment of the stochastic system, where
the adiabatic conditions $\vartheta, \theta\ll 1$ are applicable.
Then, two last equations of the system
(\ref{E}) arrive at dependencies type of Eqs.~(\ref{X1})
\begin{eqnarray}
\Sigma_q(t)=\overline\Sigma_q + \widetilde\Sigma_q~\xi(t),\quad
\zeta_q(t)=\bar\zeta_q + \tilde\zeta_q~\xi(t),
\label{Ea}
\end{eqnarray}
where deterministic and fluctuational components are determined as follows (cf. Eqs.~(\ref{X2}))
\begin{eqnarray}
\overline\Sigma_q &\equiv &\zeta^0 s^{\tau/2} d_\tau(s),\quad
\widetilde\Sigma_q\equiv\sqrt{I_\Sigma+I_\zeta s^\tau}~d_\tau(s);\nonumber\\
\label{Eb}\\
\bar\zeta_q &\equiv &\zeta^0~ d_\tau(s),\quad
\tilde\zeta_q\equiv\sqrt{I_\zeta+I_\Sigma s^\tau}d_\tau(s),
\qquad d_\tau(s)\equiv (1+s^\tau)^{-1}.\nonumber
\end{eqnarray}
So, due to the slaving principle of synergetics,
initially adiabatic noises of the complexity and energy are transformed
to multiplicative form.
On the other hand, relation between the complexity and energy
\begin{eqnarray}
\overline\Sigma_q=\sqrt{\bar\zeta_q(\zeta^0 - \bar\zeta_q)},
\label{Ec}
\end{eqnarray}
following from the dependencies (\ref{Eb}), arrives at expression
\begin{eqnarray}
T=-\left(1-{\zeta^0\over 2\bar\zeta_q}\right)^{-1}\sqrt{{\zeta^0\over\bar\zeta_q} - 1}
\label{Ed}
\end{eqnarray}
for effective temperature $T\equiv\partial\bar\zeta_q/\partial\overline\Sigma_q$.
As is depicted in Fig.~10a, so defined temperature increases monotonically with energy growth
from magnitude $T=0$ at $\bar\zeta_q=0$ to infinity at the point $\bar\zeta_q=\zeta^0/2$.
Here, the temperature $T$ breaks abruptly to negative infinity and then
increases monotonically again to initial magnitude $T=0$ at $\bar\zeta_q=\zeta^0$.
This means that inside domain $0\leq\bar\zeta_q<\zeta^0/2$ the avalanche system is dissipative
to behave in usual manner; contrary, within domain $\zeta^0/2<\bar\zeta_q\leq\zeta^0$
self-organization process evolves, so that an energy increase derives to complexity decrease,
in accordance with negative value of temperature.
At steady state, where avalanche has stationary
size $s_0=\sqrt{\zeta^0 - 1}$, the temperature takes the
stationary magnitude
\begin{eqnarray}
T_0=-~{\sqrt{\zeta^0 - 1}\over 1-\zeta^0/2}
\label{Ee}
\end{eqnarray}
being negative within supercritical domain $1\leq\zeta^0<2$.
According to Fig.~10b, the stationary temperature $T_0$ decreases monotonically
with the driven energy from the zeroth magnitude at $\zeta^0=1$ to negative infinity at $\zeta^0=2$.

Presented self-organization regime addresses to externally driven systems,
which are relevant to the usual phase transition but not to
the SOC itself.
To study the latter within above consideration, let us combine Eqs.~(\ref{Ea}), (\ref{Eb})
with the first of equations (\ref{E}) along the line,
which has been used above for obtaining Langevin equation (\ref{VI.5_17a}).
So, by analogy with Section\ \ref{sec:level4},
we arrive at stochastic equation
(\ref{9ad}), where effective force and noise intensity are given by Eqs.~(\ref{9a})
with accuracy to the replacements mentioned after Eqs.~(\ref{F}): the quantities
$s$, $\Sigma_q$, $\zeta_q$, $\tau/2$ have to be taken
instead of $u$, $v$, $S$, $a$.
Then, all results obtained in Section\ \ref{sec:level4}
can be used immediately. Particularly,
it is appeared that influence of random scattering of
avalanche size is non-essential, whereas
energy and complexity noises affect crucially.
Related picture is reflected by Fig.~8 taken in plane $I_\zeta - I_\Sigma$
formed by corresponding noise intensities of avalanche ensemble.
By this, mixed domain A+N respected to intermittency regime
is bounded by the straight line (\ref{9z}) and bell-shaped curve type of Eqs.~(\ref{ca}).
According to Fig.~9, where exponent $a$ has to be replaced by $\tau/2$,
random scattering growth of the complexity
extends the SOC domain along axis of exponents $\tau$.

We are in position now to consider the avalanche size distribution
on the basis of equations (\ref{9a}) -- (\ref{9pp}).
At arbitrary relations of noise intensities, general expressions are as follows:
\begin{eqnarray}
&P(s)={Z^{-1}\over I(s)}\exp\left\{\int\limits^s_0{f(s')\over I(s')}{\rm d}s'\right\};&\nonumber\\
&f(s)\equiv -s+\zeta^0 s^{\tau/2}d_\tau(s),&\label{Y1}\\
&I(s)\equiv I_s + \left(I_\Sigma + I_\zeta s^{\tau}\right)d^2_\tau(s),\quad
d_\tau(s)\equiv\left(1+s^\tau\right)^{-1}.&\nonumber
\end{eqnarray}
At SOC regime, the driven energy $\zeta^0=0$ and the distribution (\ref{Y1}) behaves as depicted in Fig.~11
for different noise intensities of both energy and complexity.
It is seen that the power-law dependence inherent in SOC regime
is observed only within limits $s\ll 1$ and $I_s, I_{\Sigma}\ll I_{\zeta}$ (at condition $\zeta^0=0$).
Here, the distribution (\ref{Y1}) is reduced
to the canonical form (\ref{0b}),
where second multiplier takes the form
\begin{eqnarray}
{\cal P}(s)={d_\tau^{-2}(s)\over Z}
\exp\left\{-I_{\zeta}^{-1}\int\limits^s_0{d_{\tau}^{-2}(s')\over (s')^{\tau - 1}}
{\rm d}s'\right\},\quad d_\tau(s)\equiv\left(1+s^\tau\right)^{-1}.
\label{Y2}
\end{eqnarray}
It is easily to see that deviation of this multiplier off a constant value
is estimated with term $\sim s^{2-\tau}$, whose contribution
increases remarkably with
$\tau$-decrease and growth of avalanche size
to extremely large magnitudes $s\sim 1$, i. e., with escaping SOC domain.
This confirmed by Fig.~12, where deviation $\delta\tau$ of the slope of
dependence $P(s)$, Eqs.~(\ref{Y1})
within linear domain from the theory parameter $\tau$ is depicted
as function of the parameter $\tau$ itself.
It is seen, in accordance with above estimation, that the
deviation $\delta\tau$ takes
maximal value $\delta\tau < 10^{-1}\tau$ at non-essential
magnitudes $\tau <1$ or
with noise intensity growth to enormous values $I_S\sim 10^3$.

\section{Discussion}\label{sec:level6}

Remarkable peculiarity of expression (\ref{Y2}) is that,
within limits $s\ll 1$, $I_s, I_{\Sigma}\ll I_{\zeta}$  inherent in SOC regime,  it can be rewritten
in the form
\begin{eqnarray}
{\cal P}(s)={d_{\tau}^{-2}(s)\over Z}
\exp\left\{-~{\Gamma(2-\tau)\over I_{\zeta}}
~{\cal I}^{2-\tau}_{-s}d_{\tau}^{-2}(s)\right\}
\label{K}
\end{eqnarray}
expressed in terms of standard Gamma-function $\Gamma(x)$ and
fractional integral ${\cal I}^{2-\tau}_{-s}$ of order $2-\tau$, whose definition (\ref{L})
is given in Appendix (for details, see Refs.~\cite{15}, \cite{15a}).
On the other hand, it is well-known \cite{16} that such type expressions appear
as a solution of fractional Fokker-Planck equation
\begin{eqnarray}
{\cal D}^\omega_t{\cal P}(s, t)=
{\cal D}^\varpi_{-s}\left\{s{\cal P}(s, t)
+{I_\zeta\over\Gamma\left(\varpi\right)}{\cal D}^{\varpi}_{-s}\left[d_{\tau}^{2}~{\cal P}(s, t)\right]\right\},
\label{Y3}
\end{eqnarray}
where fractional derivative ${\cal D}^{\varpi}_x$ defined by Eq.
(\ref{M}) is used to be inverted to the fractional integral
(\ref{L}).  Multiplying equation (\ref{Y3}) by term $s^{2\varpi}$,
for a $\alpha$-average
\begin{eqnarray}
|s|\equiv\langle s^\alpha\rangle^{1\over\alpha},\quad
\langle s^\alpha\rangle\equiv\int\limits^{\infty}_{-\infty}s^\alpha{\cal P}(s, t){\rm d}s,\quad
\alpha>0
\label{M1}
\end{eqnarray}
one obtains at $\alpha\equiv 2\varpi$
\begin{eqnarray}
|s|^z\sim t,\quad z={2\varpi\over\omega},
\label{M2}
\end{eqnarray}
where $z$ is dynamical exponent. This relation corresponds to
large time limit, where only diffusional
contribution is essential. Combining
expressions (\ref{K}), (\ref{M2}) and (\ref{L}) arrives at relations
$2-\tau= \varpi = z\omega/2$, which yields
\begin{eqnarray}
\tau= 2 -{z\omega\over 2}.
\label{M3}
\end{eqnarray} Comparing this
equation with the second of known relations (\ref{0a}), one obtains
\begin{eqnarray}
\omega z={4\over D}.
\label{M4}
\end{eqnarray}
Then, mean-field magnitudes $\omega = 1$ and $D=4$ are related
to dynamical exponent $z=1$
that, in accordance with definition (\ref{M2}), addresses to unusual ballistic limit of SOC regime.
On the other hand, the fractional Fokker-Planck equation (\ref{Y3})
arrives at usual diffusional regime with $z=2$ only
in artificial case, when time-derivative exponent is supposed to be
$\omega=1/2$.

Obvious reason for such discrepancy is non-consistent application of usual field relations (\ref{0a})
to Lorenz system (\ref{E}). In this system, the
role of different space directions is played by stochastic degrees
of freedom $s$, $\Sigma_q$ and $\zeta_q$, whose number is $n = 3$.
However, stochastic process evolves for every of these variables in plane spanned by given variable itself
and conjugated momentum. Moreover, multiplicative character of noise,
which is determined by exponent $a$ in expressions (\ref{9a}), reduces fractal dimension of every plane
to the value $2(1-a)$ \cite{14}.
Thus, resulting fractal dimension of phase space, where stochastic system evolves, is as follows:
\begin{eqnarray}
D = 2n(1-a),
\label{M5}
\end{eqnarray}
where one has $n =3$ for used Lorenz system.
Insertion of this dimension into expression (\ref{M4}) arrives at
the connection $\omega z=2$, which
in contrast to above obtained relation $\omega z=1$ is correct in the simplest
case $\omega=1$, $z=2$ [the latter is relevant to
single stochastic freedom degree ($n=1$) with additive noise ($a=0$)].
In general case, equations (\ref{M3}) -- (\ref{M5}) yield final result
\begin{eqnarray}
\tau= 2\left[1 - {1\over 2n(1-a)}\right].
\label{M6}
\end{eqnarray}
Respective dependencies are depicted in Figs.13a,b to show that exponent $\tau$
increases monotonically from minimum magnitude $\tau=1$ at critical number $(1-a)^{-1}$
to upper value $\tau =2$ in limit $n\to\infty$; thereby, $a$-growth shifts dependence $\tau(n)$
to large magnitudes $n$, i. e., decreases  exponent $\tau$.

It is easily to see that relation (\ref{M6}) reproduces known results of different approaches
for dimension $D$ (see Ref.~\cite{17}).
In the case related to mean-field theory, one has $\tau=3/2$ and equation (\ref{M6})
expresses number of self-consistent stochastic equations being in
need of treating SOC behavior as function of exponent of the corresponding multiplicative noise:
\begin{eqnarray}
n={2\over 1-a}.
\label{M7}
\end{eqnarray}

Thus, in accordance with Fig.~13c, self-consistent mean-field treatment
becomes possible if the number of relevant equations is more than
minimum magnitude $n_{c}=2$.
Approaches \cite{2b} -- \cite{2cc}, \cite{10}, \cite{11} represent
examples of such considerations, where noise is supposed to be
additive in character ($a=0$).
Switching multiplicative noise arrives at $a$-growth and
non-contradicting representation of SOC demands increasing
number of self-consistent equations:
for example, within the field scheme \cite{11a} related to directed percolation ($a=1/2$),
mean-field approximation is applicable for dimensions more the critical magnitude $d_c=4$;
used here and in Refs.~\cite{12}, \cite{J} Lorenz scheme ($n=3$)
addresses
to multiplicative noise characterized by the exponent $a=1/3$ (see below).

Let us focus now on relation of above exponents to nonextensivity parameter $q$
addressed to Tsallis definitions (\ref{A}), (\ref{B}) \cite{Ts}.
Relevant kinetic behavior could
be described by nonlinear Fokker-Planck equation
\begin{eqnarray}
{\cal D}^\omega_t P(s, t)=
{\cal D}^{2}_{-s}P^q(s, t),
\label{M8}
\end{eqnarray}
where measure units are chosen so that effective
diffusion coefficient disappears,
$\omega>0$, $q>0$  are relevant exponents \cite{18}, \cite{19}.
Supposing here normalized distribution function in self-similar form type of Eq.~(\ref{0b})
\begin{eqnarray}
P(s, t)=s^{-1}_c{\cal P}(x);\qquad s_c\equiv s_c(t),\quad x\equiv s/s_c,
\label{M9}
\end{eqnarray}
we obtain
\begin{eqnarray}
s_c^q\sim t^\omega,\quad {\cal P}^{q-1}\sim x.
\label{M10}
\end{eqnarray}
On the other hand, we could use the fractional Fokker-Planck equation type of Eq.~(\ref{Y3}):
\begin{eqnarray}
{\cal D}^\omega_t P(s, t)=
{\cal D}^{2\varpi}_{-s}P(s, t).
\label{M11}
\end{eqnarray}
Inserting here the solution (\ref{M9}), one finds dependencies
\begin{eqnarray}
s_c^{2\varpi}\sim t^\omega,\quad {\cal P}^{2\varpi-1}\sim x,
\label{M12}
\end{eqnarray}
whose comparison with Eqs.~(\ref{M10}) yields
\begin{eqnarray}
q=2\varpi.
\label{M13}
\end{eqnarray}
Because of the average $|s|$ in Eq.~(\ref{M1}) is reduced to the scale $s_c$ in the case of self-similar systems,
relevant dependencies (\ref{M2}), (\ref{M10}) and (\ref{M12}) get
\begin{eqnarray}
q=z\omega.
\label{M13d}
\end{eqnarray}
Combining this equality with Eqs.~(\ref{M4}), (\ref{M5})
arrives at the resulting expression
for nonextensivity parameter of the system under consideration:
\begin{eqnarray}
q={2\over n(1-a)}.
\label{M14}
\end{eqnarray}
Maximum magnitude $q=2/n$ is related
to systems with additive noise ($a=0$), which is
relevant to mean-field picture at number of governing equations $n<2$ ($n=1$, really).
Switching multiplicative noise with increasing exponent $a>0$ arrives at $q$-growth and
self-organizing system becomes nonextensive in character ($q\geq 1$)
at the boundary (\ref{M7}) of the mean-field applicability domain.
In accordance with above estimation, fractional Lorenz system is nonextensive essentially
if the exponent $a>1/3$.

It is worthwhile to remember that we have addressed above the superdiffusion process only to be
related to L\'evy flights at discrete time instants with arbitrary displacements, including infinite ones \cite{17c}.
Being related to the Fokker-Planck equation (\ref{Y3}), such type processes are characterized 
by exponents $\omega=1$ and $\varpi<1$, the first of which is constant, 
whereas the second one characterizes fractal time-sequance of the L\'evy flights 
to arrive at the dynamical exponent $z\equiv 2\varpi/\omega<2$ (see the last of equations (\ref{M2})). 
To be dependent on microscopic conditions,  
the probability distribution of displacements ${\bf x}$ reads
\begin{eqnarray}
p({\bf x})\sim x^{-(D+\gamma)}
\label{M14a}
\end{eqnarray}
that is characterized by the fractal dimension $D$ and microscopic
step exponent $\gamma$.
It is appeared that in the case of rare events, when
$\gamma<2$, the dynamical exponent $z$ is reduced to the microscopic
step exponent ($z=\gamma<2$), whereas at $\gamma\geq 2$
one has $z=2$ \cite{17b}.  

In the opposite case of subdiffusion process, a
microscopical ingredient is random L\'evy walks instead of the
discrete L\'evy flights.  This process is involved continuously in
course of the time over discrete placed traps, so that the 
exponents $\omega<1$ addresses fractal properties of this space to be dependent on microscopic conditions.
This properties produce transformation of the usual Boltzmann-Gibbs statistics in nonextensive manner \cite{Ts}.
Along this line, the subdiffusion process is
presented by the Tsallis-type distribution \cite{17a}
\begin{eqnarray}
p({\bf x})\propto\left[1+\beta(q-1) x^{2}\right]^{-~{1\over q-1}},\quad\beta={\rm const}>0,
\label{M14b}
\end{eqnarray}
where deviation $q-1$ of the nonextensive parameter is caused by fractal nature 
of the system phase space to be connected with the step
exponent $\gamma$ as follows:
\begin{eqnarray}
q=1+{2\over D+\gamma}.
\label{M14c}
\end{eqnarray}
Formal advantage of the distribution (\ref{M14b}) is that corresponding $q$-weighted average
\begin{eqnarray}
\langle{\bf x}^2\rangle_q\equiv\int {\bf x}^2 p^q({\bf x}){\rm d}^D x,
\label{M14d}
\end{eqnarray}
where the integrand varies as $x^{-(1+\gamma)}$, is converges for arbitrary
step exponents $\gamma>0$.
As a result, the law of motion of the random L\'evy walker is given by
\begin{eqnarray}
\langle{\bf x}^2\rangle_q
\sim t^\omega,\quad
\omega=\left\{
\begin{array}{ll}
q-1 &{\rm at}\quad\gamma, q<2,\\
1-(q-1){D\over 2}
&{\rm at}\quad\gamma\geq 2,~q>1.
\end{array} \right.
\label{M12a}
\end{eqnarray}
In contrast to Eq. (\ref{M12}), where the exponents
$\varpi<1$, $\omega=1$ are relevant to
the superdiffusion, here one has inverted relations $\varpi=1$, $\omega<1$.
Thus, in accordance with subdiffusion nature, the last equality (\ref{M2}) yields 
the dynamical exponent $z>2$. 

In general case $\varpi, \omega\ne 1$,
inserting Eqs. (\ref{M12a}) into the relation (\ref{M13d}) arrives at the result
\begin{eqnarray}
z=\left\{
\begin{array}{ll}
{q\over 1-{D\over 2}(q-1)}
&{\rm at}\quad 1<q\leq q_D,\\
{q\over q-1} &{\rm at}\quad q_D\leq q\leq 2,
\end{array} \right.
\label{M12b}
\end{eqnarray}
where boundary value of the nonextensivity parameter is introduced:
\begin{eqnarray}
q_D\equiv{4+D\over 2+D}.
\label{M12c}
\end{eqnarray}
To avoid a mistake, let us focus that in contrast to equalities
(\ref{0a}), (\ref{M3}), (\ref{M4}), which could be addressed to both the
real phase space and the configurational one (the latter is spanned by variables 
of governing equations), 
above obtained relations (\ref{M12a}) -- (\ref{M12c}) 
are relevant to the real phase space only. 
This is reflected by addressing the fractal dimension $D$
to the only real coordinate space in the former case, whereas in the latter it is reduced to 
the effective value (\ref{M5}). 
Along the principle line of our treatment, central role is played
by the relation (\ref{M5}) since, by analogy with studying renormalization group, 
we have considered mainly the properties of the configurational space 
but not real diffusion process.

Finally, let us address the question: Why the Lorenz system is used
but not R\"ossler one or another?
The reason can be recognized within supersymmetry field approach,
whose using shows that the Lorenz system could be generated
by the Langevin equation for order parameter
to be relevant to standard stochastic system \cite{20}.
On the other hand, one can see within microscopic consideration that
Lorenz system addresses to the simplest Hamiltonian of a bozon-fermion system \cite{14a}.
To convince in this statement, let us consider such system
with interaction $w$. Bozons are described
by creation and annihilation  operators $b_l^{+}$, $b_l$, satisfying the usual
commutation relation:  $[b_l,b_m^{+}]=\delta_{lm}$, where $l,m$ are
the site numbers.  The two-level Fermion subsystem is represented by
operators $a_{l\alpha}^{+}, a_{l\alpha}$, $\alpha=1,~ 2$, for which
the anti-commutation relation $\{a_{l\alpha},
a_{m\beta}^{+}\}=\delta_{lm}\delta_{\alpha\beta}$ is fulfilled.  The
occupation numbers $b_{\rm{\bf k}}^{+}b_{\rm{\bf k}}$ determine the
Bozon distribution within $\rm{\bf k}$-representation that corresponds
to the Fourier transform over lattice sites $l$. To represent the
Fermi subsystem we should introduce the operator $d_{l}\equiv
a_{l1}^{+}a_{l2}$ determining the polarization with respect
to the saturation over levels $\alpha=1,~2$, as well as
the occupation numbers $n_{l\alpha}\equiv
{a_{l\alpha}^{+}a_{l\alpha}}$.  As a result, behavior of the
system is defined by Dicke Hamiltonian
\begin{equation}
H=\sum_{\rm{\bf k}} \left\{(E_{1}n_{{\rm{\bf k}}1}+
E_{2}n_{\rm{\bf k}2})+\omega_{\rm{\bf k}}b_{\rm{\bf k}}^{+}b_{\rm{\bf
k}} +{{\rm i} \over 2} w({b_{\rm{\bf k}}^{+}}d_{\rm{\bf
k}}-d_{\rm{\bf k}}^{+}b_{\rm{\bf k}}) \right\},
\label{Z1}
\end{equation}
where $E_{1,2}$ are energies of the Fermi levels, $\omega_{\rm{\bf k}}$
is the Bozon dispersion law, the imaginary unit before
the interaction $w$ reflects the Hermitian property and
the Planck constant is $\hbar=1$.

The Heisenberg equations of motion corresponding to Hamiltonian
(\ref{Z1}) have the form
\begin{eqnarray}
\dot{b}_{\rm{\bf k}}&=&-{\rm
i}\omega_{\rm{\bf k}}b_{\rm{\bf k}}+{w\over 2}d_{\rm{\bf k}};\\
\label{ra}
\dot{d}_{\rm{\bf k}}&=&-{\rm i}\Delta d_{\rm{\bf k}}+{w\over 2}b_{\rm{\bf
k}}(n_{\rm{\bf k}2}-n_{\rm{\bf k}1}),\quad \Delta\equiv{E_{2}-E_{1}};\\
\label{Lb}
\dot{n}_{\rm{\bf
k}1}&=&{w\over 2}(b_{\rm{\bf k}}^{+} d_{\rm{\bf k}}+d_{\rm{\bf k}}^{+}
b_{\rm{\bf k}}),\quad
\dot{n}_{\rm{\bf
k}2}=-~{w\over 2}(b_{\rm{\bf k}}^{+} d_{\rm{\bf k}}+d_{\rm{\bf k}}^{+}
b_{\rm{\bf k}}).
\label{Ld}
\end{eqnarray}
\noindent
At condition of resonance, the first terms in the right hand
sides of  equations (\ref{ra}), (\ref{Lb}) containing frequencies
$\omega_{\rm{\bf k}}$ and $\Delta$ may be suppressed by
introducing the multipliers $\exp(-{\rm i}\omega_{\rm{\bf k}} t)$ and
$\exp(-{\rm i}\Delta t)$ for the time dependencies $b_{\rm{\bf
k}}(t)$, $d_{\rm{\bf k}}(t)$, respectively. On the other hand, to
take into account the dissipation, these frequencies acquire  imaginary
additions $-{\rm i}/\tau_{x}$, $-{\rm i}/\tau_{y}$ characterized
by relaxation times $\tau_{x}$, $\tau_{y}$ (here the conditions
${\rm Im}~\omega_{\rm{\bf k}}<0$, ${\rm Im}~\Delta<0$ reflect the
causality principle). As a result, equations (\ref{ra}), (\ref{Lb})
get the dissipative terms $-b_{\rm{\bf k}}/{\tau_{x}}$,
$-d_{\rm{\bf k}}/{\tau_y}$, where $\tau_{x}$ is the relaxation
time of Bozon distribution and $\tau_{y}$ is the Fermion
polarization time.  Obviously, one can suppose that the dissipation influences
onto the Fermi levels occupancies $n_{\rm{\bf k}\alpha}(t)$ also.
However, since the stationary values $n_{\rm{\bf k}\alpha}^0\not=0$
(in case of external drive $n_{\rm{\bf k}2}^0>n_{\rm{\bf
k}1}^0$) the dissipative terms in Eqs.~(\ref{Ld}) take
much complicated form: $\ -(n_{\rm{\bf k}\alpha}-n^0_{\rm{\bf
k}\alpha})/\tau_S$, where $\tau_S$ is the relaxation
time of the Fermion distribution over levels $\alpha=1,~2$.

Now, let us introduce the macroscopic quantities
\begin{eqnarray}
&& u_{\bf k} \equiv \langle b_{\bf k}^{+}\rangle=
\langle b_{\bf k}\rangle, \qquad
v_{\bf k} \equiv \langle d_{\bf
k}\rangle=\langle d_{\bf k}^{+}\rangle, \nonumber\\
&& S_{\bf k} \equiv \langle n_{{\bf k}2}-n_{{\bf k}1}\rangle,\qquad
S^0_{\bf k} \equiv \langle n_{{\bf k}2}^0-n_{{\bf k}1}^0
\rangle,
\label{a3}
\end{eqnarray}
where the angular brackets mean thermodynamical averaging.
Then, neglecting the correlation in distribution of particles over quantum
states and omitting dependence on the wave vector ${\rm{\bf k}}$,
the Heisenberg equations (\ref{ra}) -- (\ref{Ld}), being complemented by
dissipative terms, result in the initial Lorenz system
(\ref{2}), (\ref{4}), (\ref{5}),
whose parameters take magnitudes $a=1$, $\tau_{x}=2/w$, $D=(2w)^{-1}$.
Respectively, the dimensionless variables in the system (\ref{6a}) -- (\ref{6c})
are as follows: $u=w(\tau_{y}\tau_{S})^{1/2}\dot{x}$, $v=w(\tau_{y}\tau_{S})^{1/2}\dot{y}$
and $S=(w\tau_{y}/2)y'$.

As a result, we arrive at the conclusion that Lorenz system is relevant
microscopically to
the simplest bozon-fermion system defined by Dicke Hamiltonian (\ref{Z1}).
At first glance, it could have been shown that one can write corresponding expression
for macroscopic (effective) Hamiltonian being
a synergetic potential,
whose dependence on the freedom degrees $u$, $v$, $S$
could have generated the Lorenz system.
But such dependence is appeared to be absent because of the effective
Hamiltonian
has to take into account quite different commutation rules related to
different freedom degrees. Obvious advantage of above mentioned
supersymmetry theory \cite{20} and stated microscopic approach
is that we have explicit possibility to account
such difference. Generally, this situation is relevant to
known problem of description of systems with intermediate
statistics (see Ref.~\cite{21}).

\section{Acknowledgments}\label{sec:level8}

We would like to dedicate this paper to the memory of our friend and
colleague Dr.~Evgeny Toropov, whose activity had prompted this work.

\section{Appendix}\label{sec:level7}
\def\theequation{{A}.\arabic{equation}} \setcounter{equation}{0}

Here the basic properties of fractional integral and derivative,
as well as Jackson  derivative are quoted for convenience.
The integral of fractional order $\varpi$ is defined by equality \cite{15}, \cite{15a}
\begin{eqnarray}
{\cal I}^{\varpi}_{x}f(x)\equiv
{1\over\Gamma\left(\varpi\right)}
\int\limits^x_0{f(x')\over(x-x')^{1-\varpi}}{\rm d}x',\quad
\varpi>0,
\label{L}
\end{eqnarray}
where $f(x)$ is arbitrary function, $\Gamma(x)$ is standard Gamma-function.
To be inverted to the fractional integral, relevant derivative
${\cal D}^{\varpi}_x\equiv{\cal I}^{-\varpi}_{x}$ of order $\varpi>0$
is determined as follows:
\begin{eqnarray}
{\cal D}^{\varpi}_x f(x)\equiv
{1\over\Gamma\left(-\varpi\right)}
\int\limits^x_0{f(x')\over(x-x')^{1+\varpi}}{\rm d}x'.
\label{M}
\end{eqnarray}
At $0<\varpi<1$ it is convenient to use equation
\begin{eqnarray}
{\cal D}^{\varpi}_x f(x)\equiv
{\varpi\over\Gamma\left(1-\varpi\right)}
\int\limits^x_0{f(x)-f(x')\over(x-x')^{1+\varpi}}{\rm d}x',
\label{N}
\end{eqnarray}
where we take into account known equality $x\Gamma(x)=\Gamma(x+1)$ for $x\equiv -\varpi$.

Let us introduce now Jackson $q$-derivative, whose advantage for analyzing self-similar system
is that this derivative determines
the rate of a function $f(x)$ variation with respect to dilatation $q\ne 1$,
but not to the shift ${\rm d}x\to 0$, as in usual case $q=1$.
According to such definition, Jackson $q$-derivative reads:
\begin{eqnarray}
{\cal D}_q f(x)\equiv
{f(qx)-f(x)\over q-1},\quad q>0.
\label{O}
\end{eqnarray}
For important case of homogeneous function, being subjected to condition
\begin{eqnarray}
f(qx)\equiv q^{\alpha}f(x),
\label{P}
\end{eqnarray}
where $q>0$ is a dilatation parameter and $\alpha>0$ is an exponent,
the Jackson $q$-derivative is reduced to Jackson $q$-number:
\begin{eqnarray}
{\cal D}_qf(x)= [\alpha]_qf(x),\quad
[\alpha]_q\equiv\frac{\alpha^q - 1}{q-1}.
\label{Q}
\end{eqnarray}
It is easily to see that the value $[\alpha]_q\to\alpha$ in
the limit $q\to 1$ and varies as $q^{\alpha-1}$ at $q\to \infty$.
On the other hand, the Tsallis
$q$-logarithmic function
\begin{eqnarray}
\ln_q x\equiv{x^{q-1}-1\over q-1}
\label{Q1}
\end{eqnarray}
can be represented in the form of the  Jackson $q$-number (\ref{Q})
with the index
$\alpha=(q-1)(\ln x/\ln q)$.  Accompanied by Eqs.~(\ref{Q}) this relation
and obvious equality
\begin{equation} \ln_q(xy)=\ln_q x +
\ln_q y + (q-1)(\ln_q x)(\ln_q y)
\label{Q2}
\end{equation}
lead to important rule for the Jackson derivative:
\begin{equation} {\cal
D}_q \left[f(x)g(x)\right]=\left[{\cal D}_q
f(x)\right]g(x)+f(x)\left[{\cal D}_q g(x)\right] + (q-1)\left[{\cal
D}_q f(x)\right]\left[{\cal D}_q g(x)\right].
\label{Q3}
\end{equation}

Generalizing Eqs.~(\ref{N}), (\ref{O}), let us introduce finally a fractional $\varpi$-derivative:
\begin{eqnarray}
{\cal D}^{\varpi}f(x)\equiv
{\varpi x^{-\varpi}\over\Gamma\left(1-\varpi\right)}
\int\limits^1_0{f(x)-f(qx) \over(1-q)^{1+\varpi}}{\rm d}q,\quad
0<\varpi<1.
\label{R}
\end{eqnarray}
In the case of self-similar system,
the function $f(x)$ is homogeneous to be subjected to condition (\ref{P}).
Then, the definition (\ref{R}) is simplified:
\begin{eqnarray}
{\cal D}^{\varpi}f(x)\equiv
\left\{\alpha\right\}_{\varpi} x^{-\varpi}f(x),\quad
0<\varpi<1
\label{S}
\end{eqnarray}
to be reduced to fractional $\varpi$-number
\begin{eqnarray}
\left\{\alpha\right\}_{\varpi}\equiv
{\varpi\over\Gamma\left(1-\varpi\right)}
\int\limits^1_0{1- q^{\alpha} \over(1-q)^{1+\varpi}}{\rm d}q,\quad
0<\alpha,\quad 0<\varpi<1.
\label{T}
\end{eqnarray}
Being the combination of $\Gamma$-functions:
\begin{eqnarray}
\left\{\alpha\right\}_{\varpi}=
{\Gamma(\alpha + 1)\over\Gamma\left(\alpha+1-\varpi\right)}-
{1\over\Gamma\left(1-\varpi\right)},
\label{T1}
\end{eqnarray}
this number increases monotonically with growth of both
exponents $\alpha$ and $\varpi$ taking zeroth
magnitude on the axes $\varpi=0$, $\alpha=0$ and
characteristic values 
$\left\{{1\over 2}\right\}_{1\over 2}={\sqrt{\pi}\over 2}-{1\over\sqrt{\pi}}\simeq 0.322$, 
$\left\{1\right\}_1=1$.
Such behavior is characterized by the particular dependencies
\begin{eqnarray}
\left\{\alpha\right\}_{\varpi}=\left\{
\begin{array}{ll}
\alpha &{\rm at}\quad\varpi=1,\\
\Gamma(1+\alpha)-{1\over\Gamma(1-\alpha)}
&{\rm at}\quad\varpi=\alpha,\\
{\varpi\over\Gamma\left(2-\varpi\right)}
&{\rm at}\quad\alpha=1.
\end{array} \right.
\end{eqnarray}
Thus, if $q$-number (\ref{Q}) related to Jackson derivative (\ref{O})
tends to exponent $\alpha$ in the limit $q\to 1$, $\varpi$-number (\ref{T})
corresponding to fractional integral (\ref{R})
is reduced to factor $\alpha$ at $\varpi=1$.

\newpage

\begin{center}
{\bf FIGURE CAPTIONS}
\end{center}

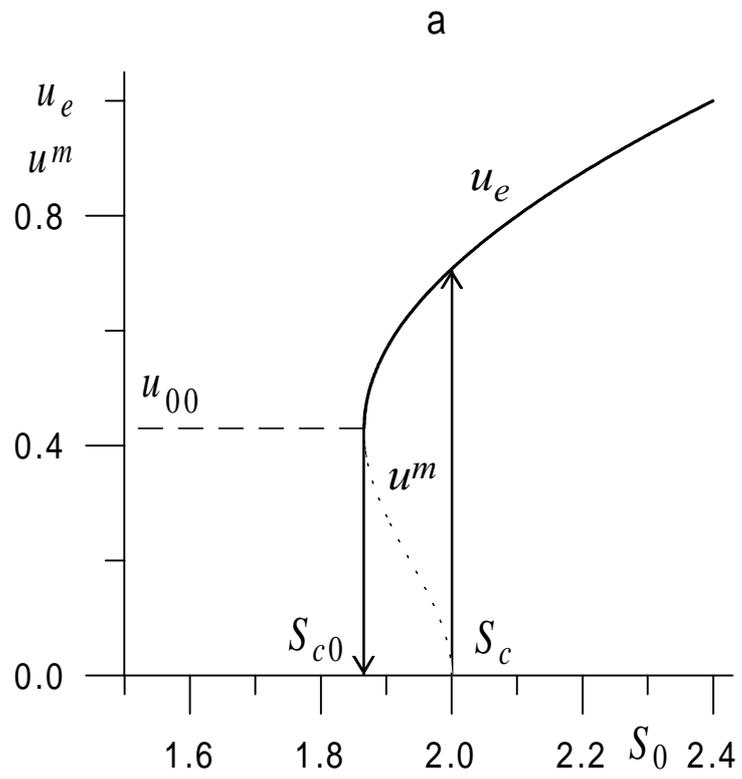
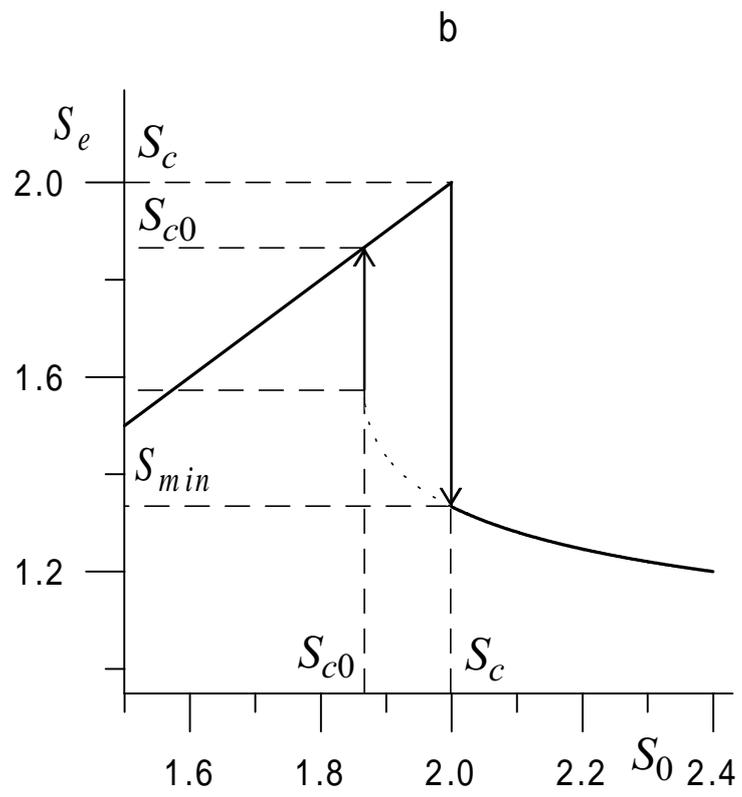
Fig.~1. The $S_{0}$-dependencies of (a) the velocities $u_{e}$,
$u^{m}$, and (b) the equilibrium slope $S_{e}$. The arrows indicate
the hysteresis loop.

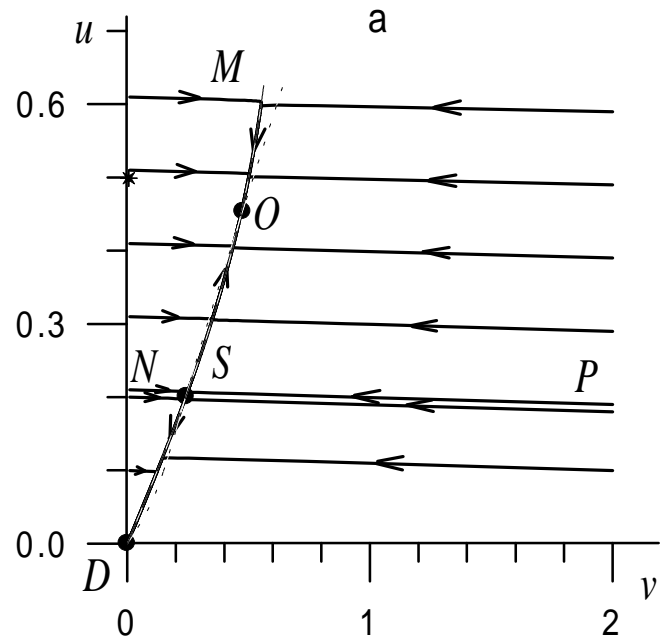
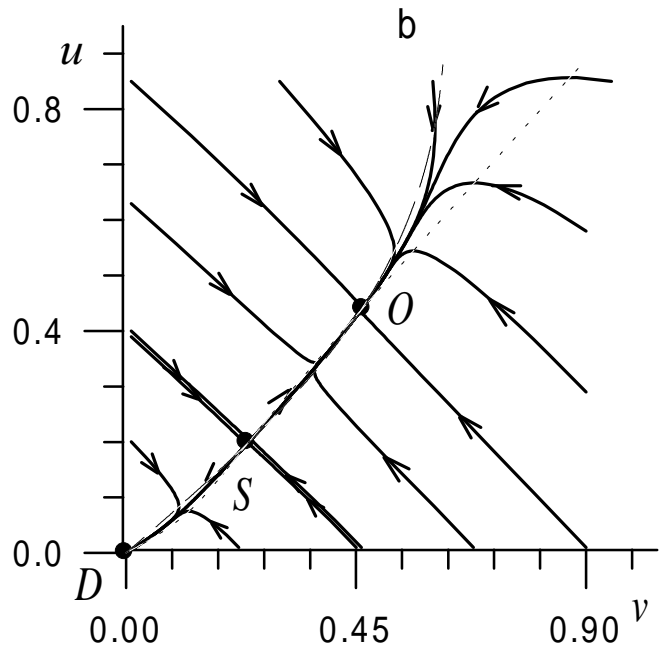
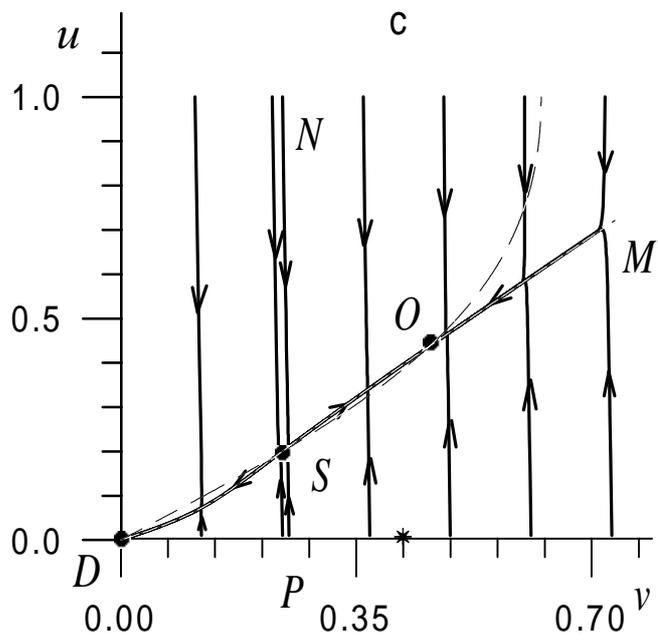
Fig.~2. Phase portraits in the $v-u$ plane at $m=1, u_0=0.1, S_{0}=1.25S_{c}$:
(a) $\epsilon=10^{-2}$; (b) $\epsilon=1$; (c) $\epsilon=10^{2}$.

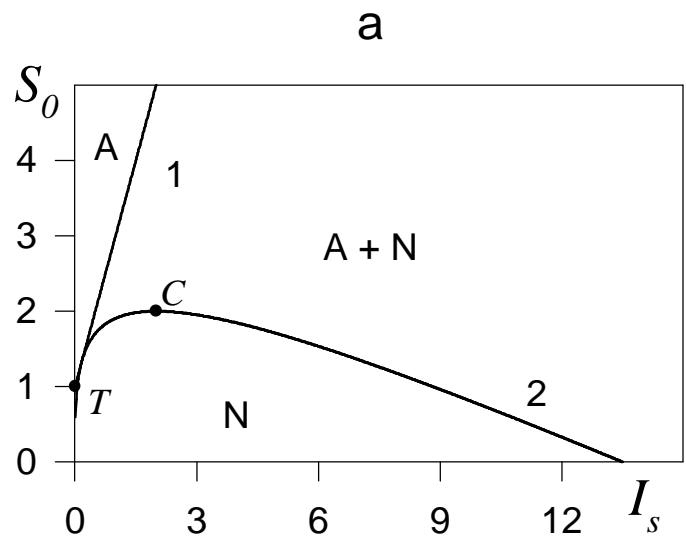
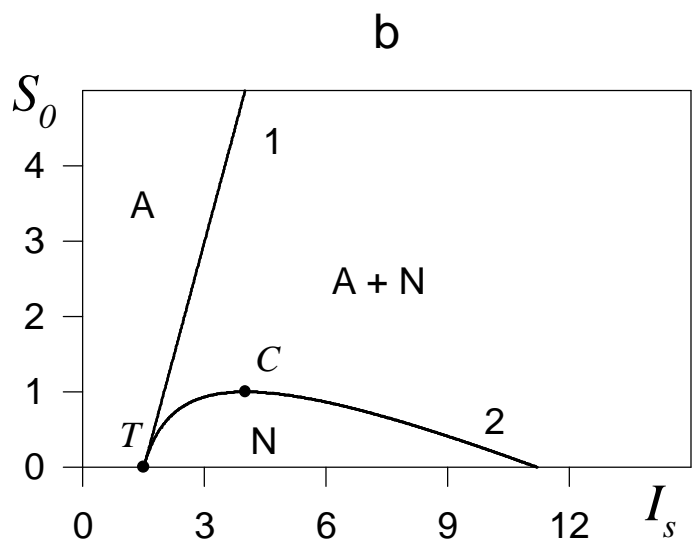
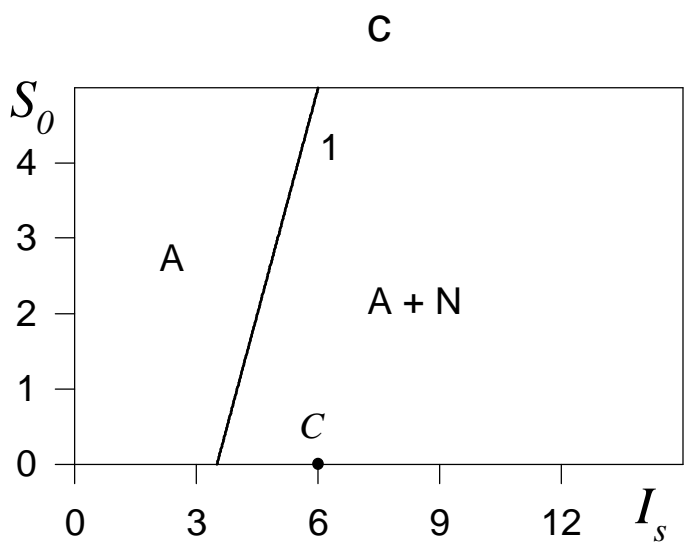
Fig.~3. Phase diagrams at fixed values $I_v$:
(a) $I_v=0$; (b) $I_v=1$; (c) $I_v=2$. Curves 1 and 2 define the
boundary of stability of avalanche (A) and non-avalanche (N) phases.

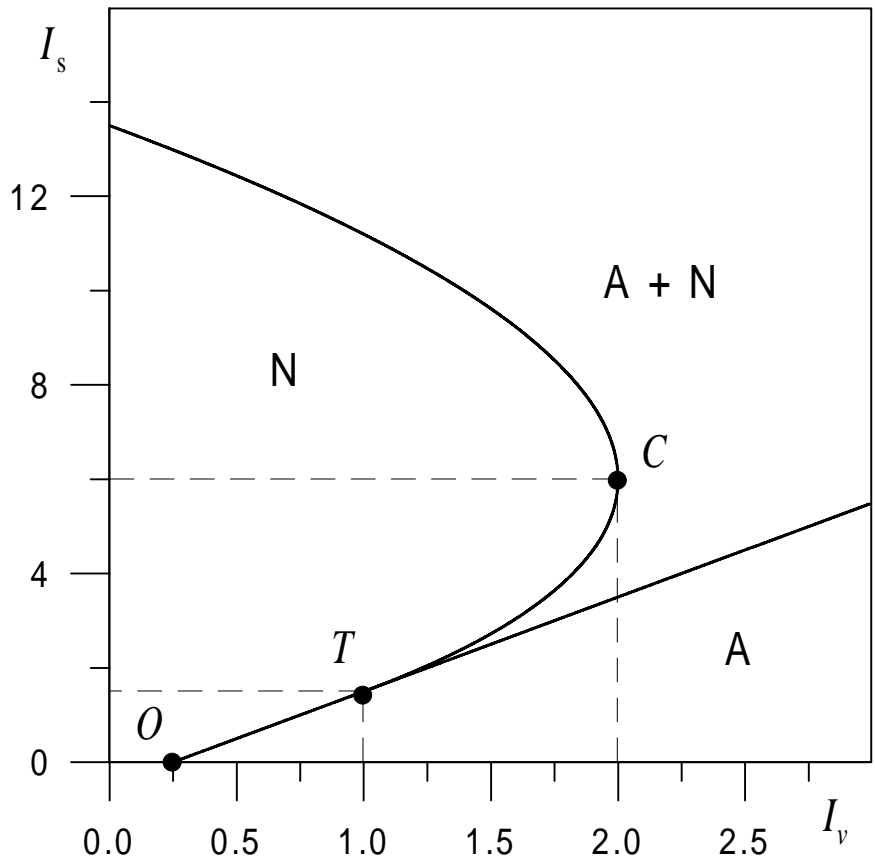
Fig.~4. Phase diagram for system with $S_0=0$ and $I_s, ~I_v\ne 0$
($D$ -- disordering point; $T$ -- tricritical
point; $C$ -- critical point).

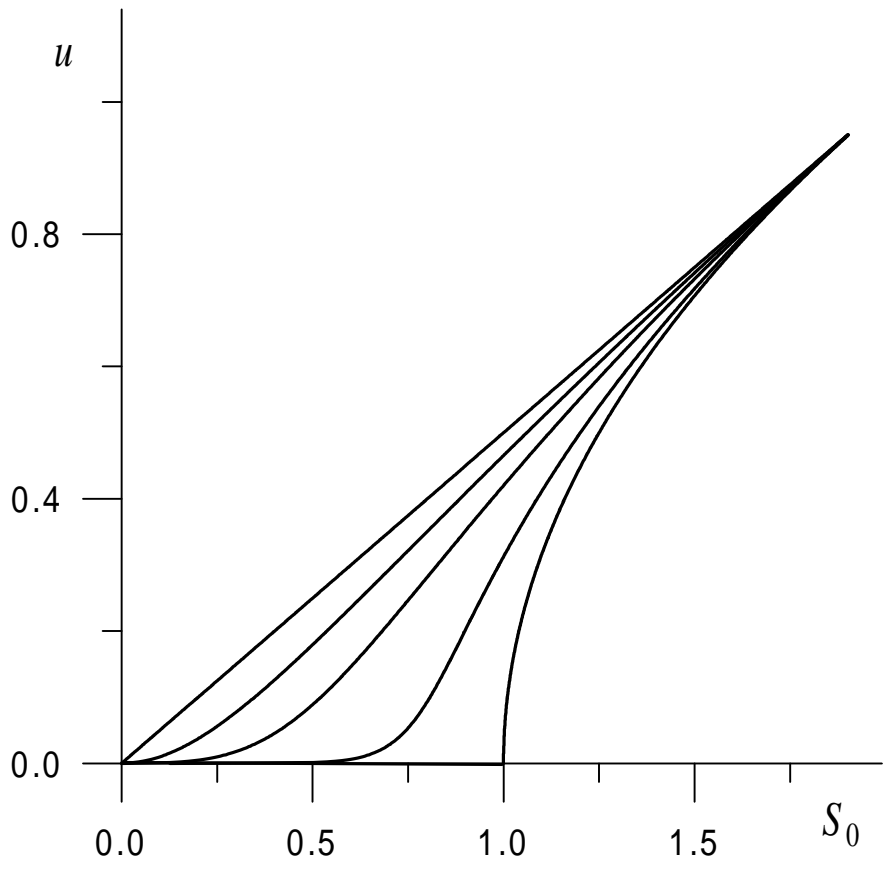
Fig.~5. The $S_{0}$-dependence of the steady-state velocity $u$  at
 $a  = 0, ~0.5, ~0.7, ~0.9, ~1.0$ from top to bottom.

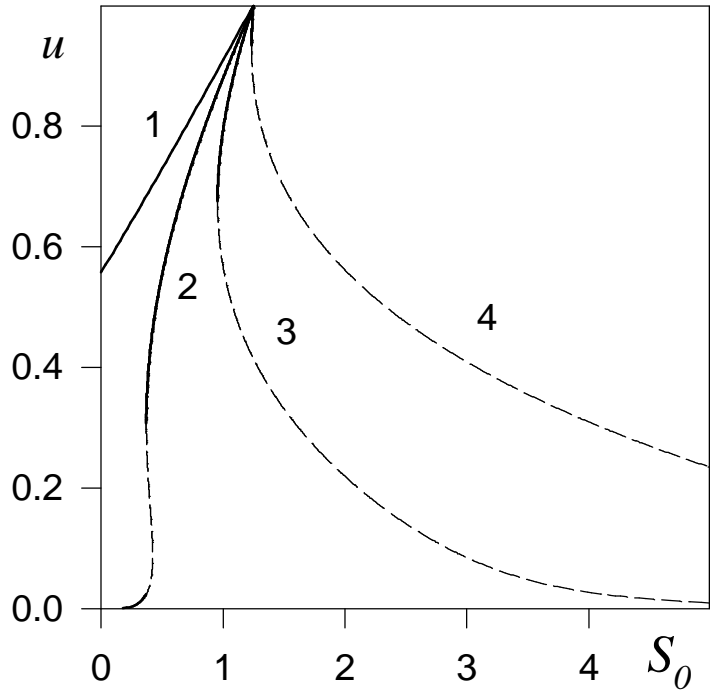
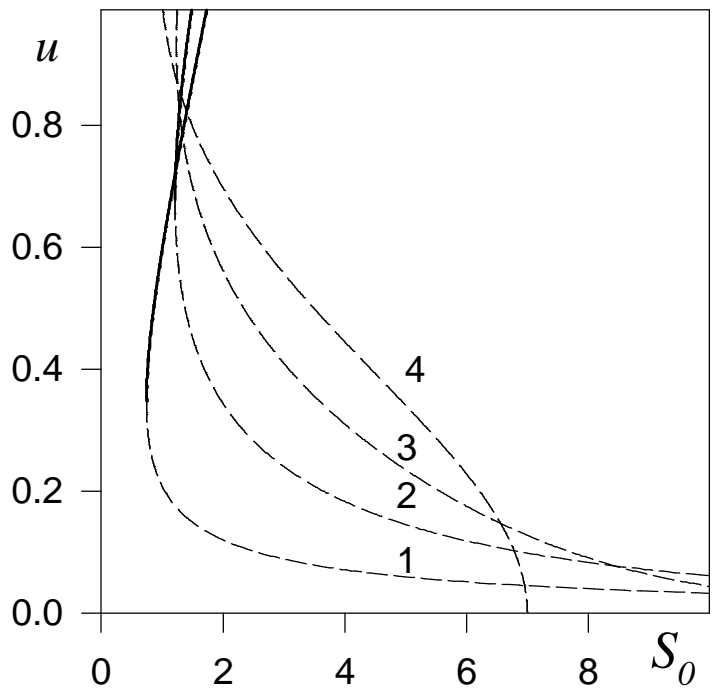
Fig.~6. The $S_{0}$-dependence of the steady-state velocity $u$:
(a) at $a  = 0.75,~I_v = 1$ (curves 1 -- 4 address to $I_S=1, ~2, ~3, ~5$); (b) at
$I_v = 1,~ I_S = 5$ (curves 1 -- 4 address to $a = 0.25, ~0.5, ~0.75, ~1.0$.

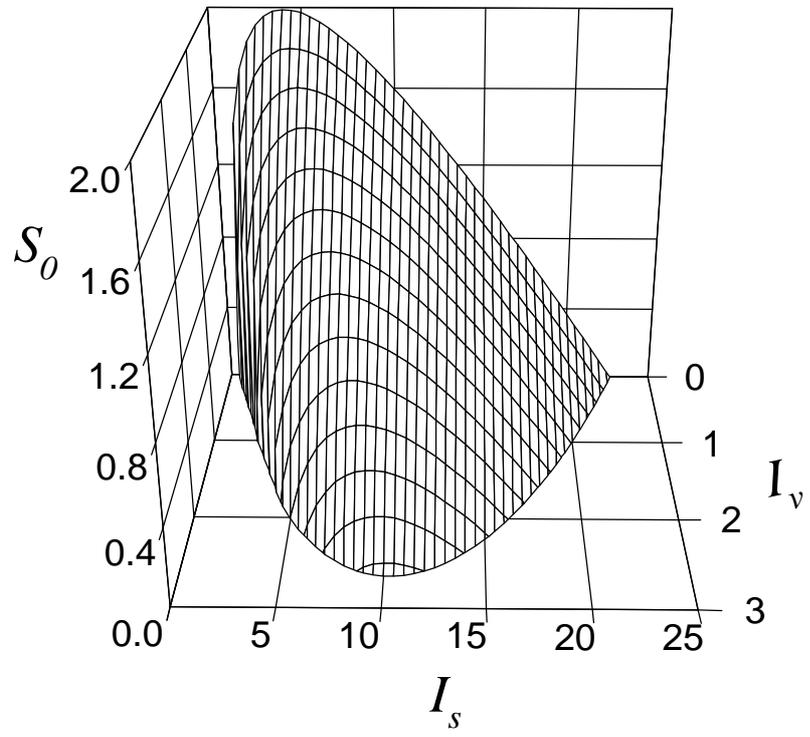
Fig.~7. Three-dimensional phase diagram (the non-avalanche domain
is located under surface).

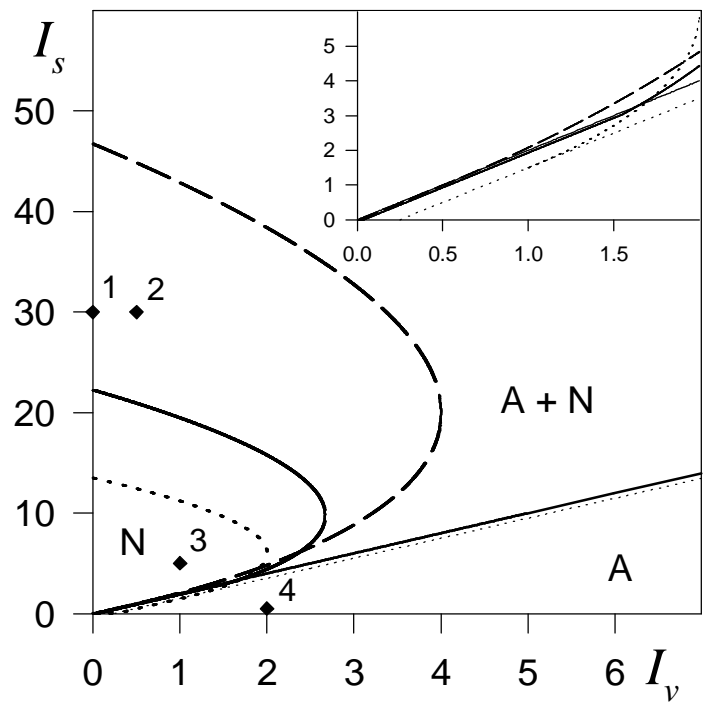
Fig.~8. Phase diagram for system with $S_0=0$ and $I_s, ~I_v\ne 0$
at $a  = 0.5, ~0.75, ~1.0$ (dotted, solid and dashed curves, respectively).
Diamonds address to curves 1 -- 4 in Fig.~11.

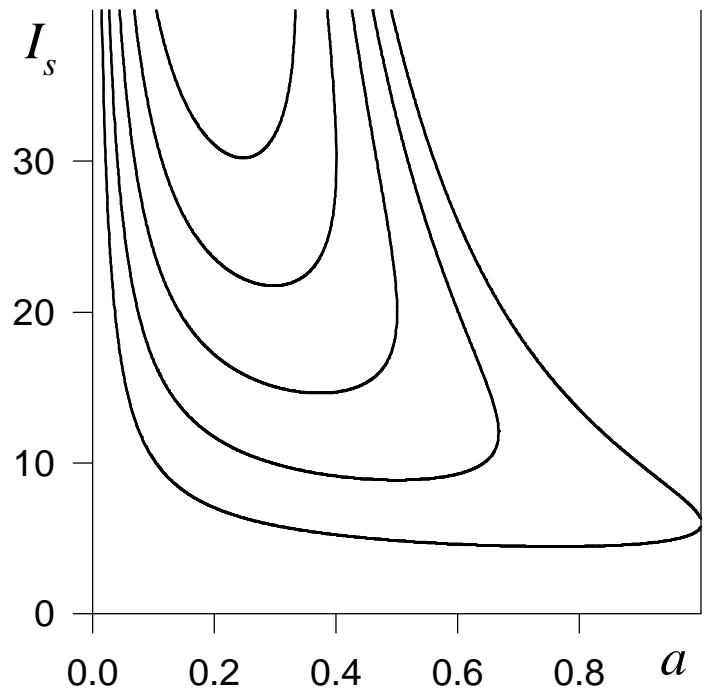
Fig.~9. Phase diagram in the $I_S - a $ plane at $I_v=2, 3, 4, 5, 6$
from bottom to top (the non-avalanche domain
is located inside the curves).

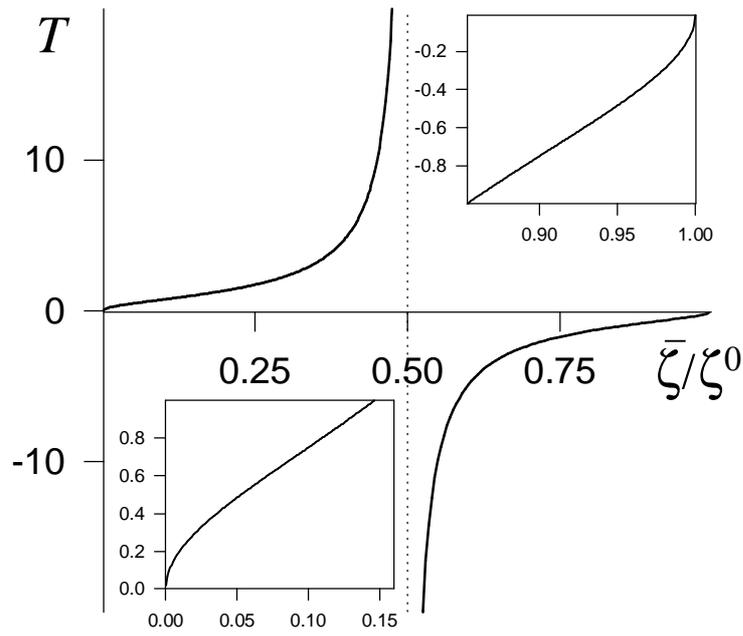
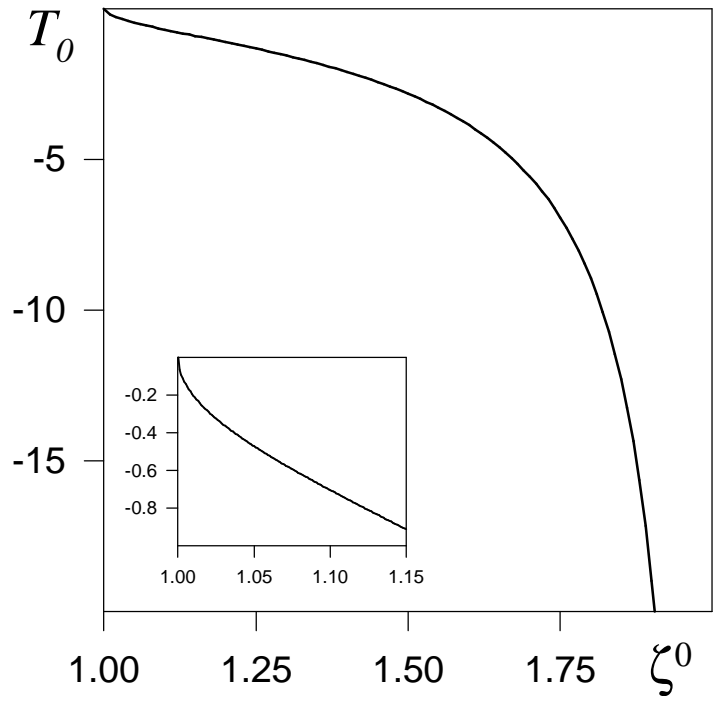
Fig.~10. The energy dependences of the avalanche ensemble temperatures:
(a) nonstationary magnitude $T$ versus ratio
$\bar\zeta_q/\zeta^0$; (b) stationary temperature $T_0$ versus  $\zeta^0$.

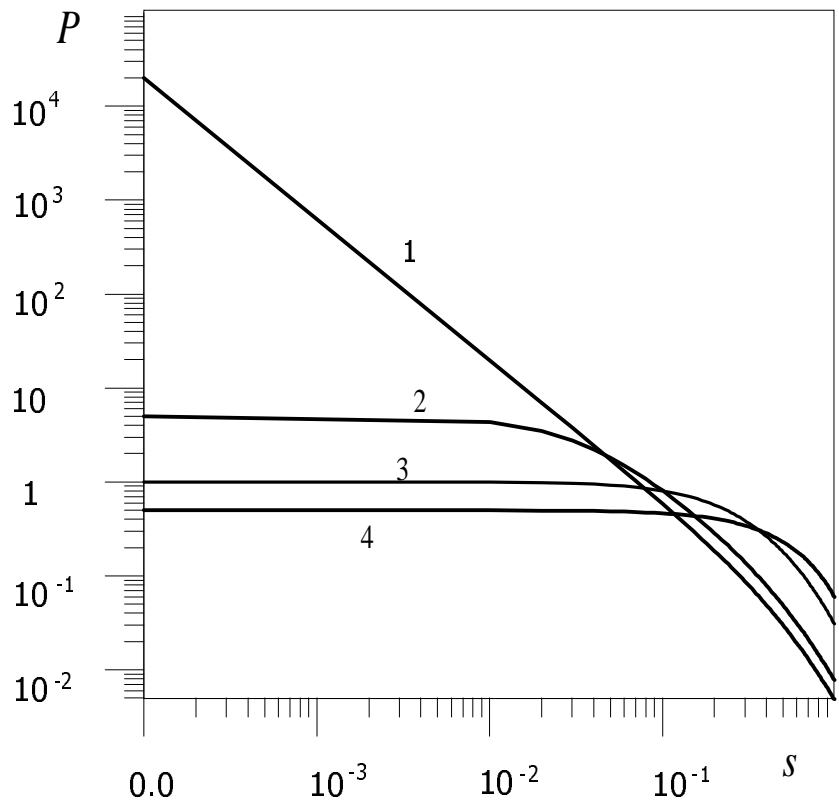
Fig.~11. Distribution function (\ref{Y1}) at $\tau=1.5$ and regimes pointed out by diamonds in Fig.~8:
1)  $I_v = 0,~ I_S = 50$ (SOC); 2)  $I_v = 0.5,~ I_S = 30$ (A+N); 3)  $I_v = 1,~ I_S = 5$ (N);
4)  $I_v = 2,~ I_S = 0.5$ (A).

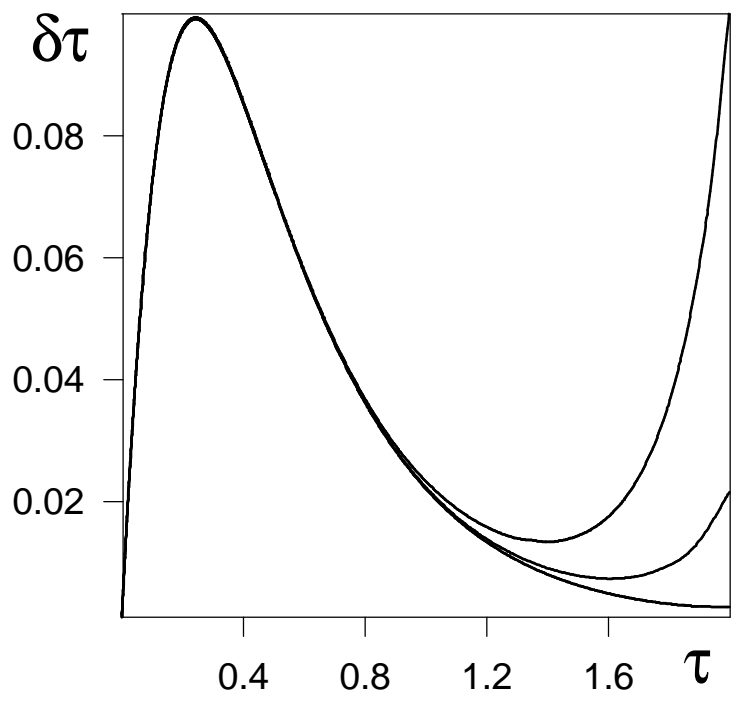
Fig.~12. Deviation $\delta\tau$ of the linear slope of curve 1 depicted in Fig.~11
from parameter $\tau$ versus the exponent $\tau$ itself
($I_S = 10, 50, 10^3$ from bottom to top).

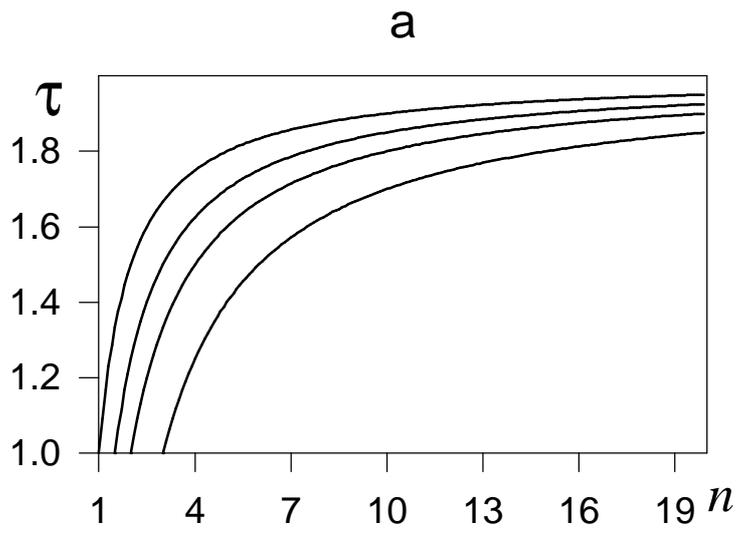
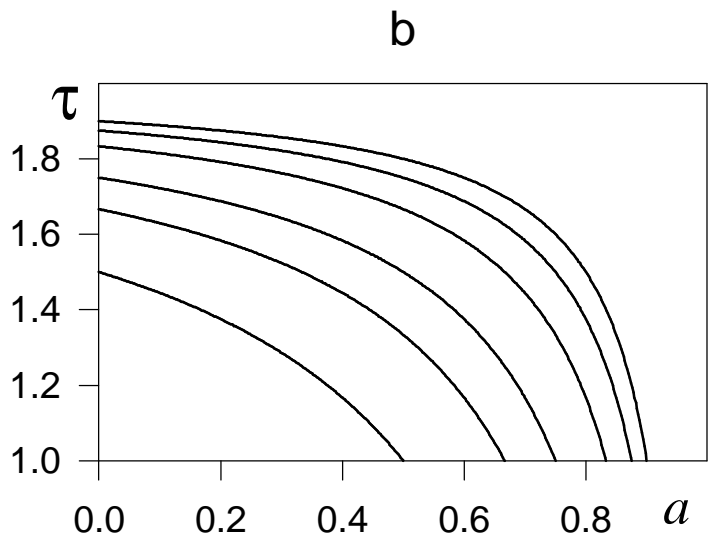
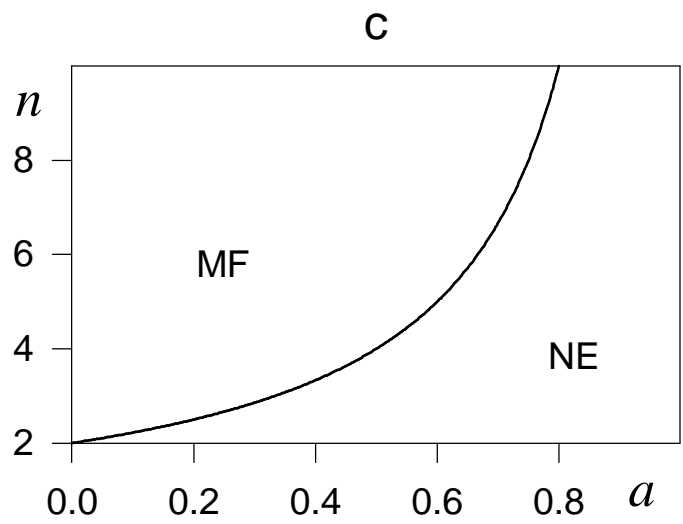
Fig.~13. Dependences of exponent $\tau$:
(a) on equations number $n$ ($a=0, {1\over 3}, {1\over 2}, {2\over 3}$ from  top to bottom);
(b) on exponent $a$ ($n=2, 3, 4, 6, 8, 10$ from bottom to top);
(c) phase diagram for mean-field and nonextensivity domains.


\end{document}